\documentclass[aps,prmaterials,preprint,floatfix]{revtex4-1}

\usepackage[utf8]{inputenc}
\usepackage{array}
\usepackage{graphicx}
\usepackage{amsmath}
\usepackage{amsfonts}
\usepackage{color}
\usepackage{dcolumn}
\usepackage{bm}

\begin{document}

\title{Topological Properties of Bulk and Bilayer 2M WS$_2$: A First-Principles Study}
\author{Nesta Benno Joseph}
\affiliation{Solid State and Structural Chemistry Unit, Indian Institute of Science, Bangalore 560012, India.}
\author{Awadhesh Narayan}
\email{awadhesh@iisc.ac.in}
\affiliation{Solid State and Structural Chemistry Unit, Indian Institute of Science, Bangalore 560012, India.}

\date{\today}

\begin{abstract}
Recently discovered 2M phase of bulk WS$_2$ was observed to exhibit superconductivity with a critical temperature of 8.8 K, the highest reported among superconducting transition metal dichalcogenides. 
Also predicted to support protected surface states, it could be a potential topological superconductor.
In the present study, we perform a detailed first-principles analysis of bulk and bilayer 2M WS$_2$.   
We report a comprehensive investigation of the bulk phase, comparing structural and electronic properties obtained from different exchange correlation functionals to the experimentally reported values. 
By calculation of the $Z_2$ invariant and surface states, we give support for its non-trivial band nature. 
Based on the insights gained from the analysis of the bulk phase, we predict bilayer 2M WS$_2$ as a new two-dimensional topological material.
We demonstrate its dynamical stability from first-principles phonon computations and present its electronic properties, highlighting the band inversions between the W $d$ and S $p$ states.
By means of $Z_2$ invariant computations and a calculation of the edge states, we show that bilayer 2M WS$_2$ exhibits protected, robust edge states. The broken inversion symmetry in this newly proposed bilayer also leads to the presence of Berry curvature dipole and resulting non-linear responses. We compute the Berry curvature distribution and the dipole as a function of Fermi energy. We propose that BCD signals, which are absent in the centrosymmetric bulk 2M WS$_2$, can be signatures of the bilayer.
We hope our predictions lead to the experimental realization of this as-yet-undiscovered two-dimensional topological material.
\end{abstract}

\maketitle


\section{\label{sec:intro}Introduction}

Transition Metal Dichalcogenides (TMDCs) have become the focus of extensive research over the last decade due to their layered nature and a wide range of interesting mechanical, electronic, structural, optical and chemical properties~\cite{2D_TMD_natrev2017,2012PhLA..376.1166Y,photocatalytic_tmdc}.
Depending on the metal coordination to the chalcogen layers, TMDCs exist in different phases -- H phase (trigonal prismatic coordination), T phase (octahedral coordination) and T' phase (distorted octahedral coordination).
In conjunction with the coordination (H, T, T'), composition (metal and chalcogen) and the crystal structure and stacking, these materials exhibit a diverse variety of electronic character, ranging from insulating to superconducting~\cite{chhowalla_tmdc_2013}. \\

H phases of NbX$_2$ and TaX$_2$ (X = S, Se) are well known for their intrinsic superconducting nature~\cite{klemm_2012} and NbSe$_2$ with a high $T_c$ value of 7.3 K~\cite{REVOLINSKY19651029} is also  widely studied due to the co-existence of both a charge density wave state and superconductivity~\cite{ugeda2016characterization,Lian_nbse2,PhysRevB.99.161119}. 
Until recently, this was the highest $T_c$ for intrinsic semiconductors reported among the TMDC family.
Apart from intrinsic superconductivity, many TMDCs exhibit unconventional superconductivity as a result of chemical doping, pressure, and electrostatic and field effect gating.
For example, alkali metal intercalated MoS$_2$ is known to show superconductivity with a maximum $T_c$ of 6.9 K~\cite{1973JChPh..58..697S,WOOLLAM1977289}.
In the case of 1T-TaS$_2$, a superconducting phase is retained over a pressure range of 3-25 GPa~\cite{sipos_Tas2_2008}.
Ising superconductivity with a maximum $T_c$ as high as 10.8 K, was observed in MoS$_2$ thin films as a result of application of a gate voltage~\cite{Ye1193}. 
Such Ising superconductors are interesting because they are possible platforms to create Majorana fermions. \\

These layered materials gathered more interest recently when topological phases were discovered among their heavier group members. 
WTe$_2$ and MoTe$_2$ were predicted~\cite{soluyanov_2015,PhysRevB.92.161107} and later experimentally verified~\cite{PhysRevLett.113.216601,wte2_mag_resistance,keum_mote2_2015} to be type-II Weyl semimetals in their bulk T$_d$ form, WTe$_2$ being the first ever material to be identified in this novel class~\cite{soluyanov_2015}.
Also, the 1T' phase of monolayer  MX$_2$ compounds are quantum spin Hall (QSH) insulators with edge states topologically protected from back scattering~\cite{Qian1344,tang2017quantum,fei2017edge,wu2018observation}. 
Coupled with strong spin orbit coupling (SOC) interactions present in the transition metal based systems, these materials are expected to be more suitable for technological applications than other QSH insulators discovered. 
The combination of strong Ising SOC, superconductivity and topological properties in TMDCs makes them very feasible candidates for the realisation of topological superconductors that can support and sustain Majorana fermions~\cite{PhysRevB.93.180501}. \\

A very recent addition to the list of different TMDCs is a new monoclinic phase named the 2M phase, composed of 1T' monolayers.
Each monolayer is then stacked in a completely different way compared to all previously known structures of TMDCs.
The recently discovered 2M WS$_2$~\cite{fang2019discovery} and 2M WSe$_2$~\cite{Fang_2020} are the first among this class.
This structure exhibits very interesting novel properties. 
Intrinsic superconductivity was measured in this phase, with a transition temperature of 8.8 K, the highest among known superconductors of the TMDC family at ambient pressure~\cite{guguchia2019nodeless,fang2019discovery,yuan2019evidence}. 
Further, this phase is predicted to be topologically non-trivial with a single Dirac cone at the Brillouin zone center, making it the first potential topological superconductor reported among TMDCs ~\cite{fang2019discovery,PhysRevB.102.024523}. \\

In this study, we investigate the properties of bulk and bilayer 2M WS$_2$ using first principles density functional calculations. We begin with a detailed study of the bulk 2M WS$_2$, systematically exploring the structural properties obtained from different types of exchange correlation functionals and their comparison to the experimentally reported values. We then present the electronic properties of the bulk phase comparing the effects of spin-orbit coupling. Next, we analyze the topological nature of the bulk phase, including a computation of the topological invariants and the surface states. We then predict bilayer 2M WS$_2$ as a new two-dimensional TMDC, which features topological band inversions. We demonstrate the dynamical stability of this new structure by means of \textit{ab initio} phonon calculations. We report the computed phonon modes and their symmetries, which may assist in experimental identification of this phase. We then present the electronic properties of this new system, highlighting the band inversions between W $d$ states and S $p$ states near the Brillouin zone center. We show that bilayer 2M WS$_2$ exhibits a non-trivial topology, by means of the topological invariant computations and explicit calculation of the edge states. We hope that our predictions lead to the experimental realization of this new two-dimensional topological material.


\section{\label{sec:methods} Computational Details}

First-principles calculations were carried out based on density functional theory (DFT) framework as implemented in the {\sc quantum espresso} code~\cite{QE-2017,QE-2009}. Taking the experimental lattice parameters as obtained by Fang \textit{et.al}~\cite{fang2019discovery}, the bulk structures of 2M WS$_2$ were relaxed using both local density approximation (LDA)~\cite{LDA} and generalized gradient approximation as given by Perdew-Burke-Ernzerhof (GGA-PBE)~\cite{perdew1996generalized} for the exchange-correlation functional.
In addition, another set of relaxations employing GGA-PBE along with semi-empirical Grimme's DFT-D2~\cite{DFT-D2} correction for the van der Waals forces was also carried out.
A kinetic energy cut-off of $750$ eV was used and all relaxations were performed until forces on each of the atoms were less than at least $10^{-4}$ eV/\AA.
The Brillouin zone was sampled over a uniform $\Gamma$-centered $k$-mesh of $2\times8\times6$ for the bulk and $1\times8\times6$ for the bilayer. Note that $a$ is the stacking direction of the layers. 
Starting with the experimental bulk 2M WS$_2$ structure, an isolated bilayer was constructed by incorporating vacuum along the $a$ axis.
The system was modelled with at least 15 \AA~vacuum to avoid any spurious interaction between the periodic images.
The structure was further allowed to relax until the forces on each atoms were less than $10^{-4}$ eV/\AA . \\

Spin-orbit coupling (SOC) was included in all calculations, as W is a heavy element. As we will see, the SOC effects can be important to describe the electronic properties in this system.
The phonon computations were performed using the {\sc phonopy} package~\cite{phonopy} interfaced with {\sc quantum espresso}. {\sc phonopy} constructs a supercell for the system and employs frozen phonon (finite displacement) approach to calculate the forces between the atoms.
The force constant matrix thus obtained is further used to calculate the normal modes of the system.
A $1\times4\times4$ supercell was constructed and an atomic displacement distance of 0.01 \AA~was considered during the calculation of normal modes associated with the bilayer system.
To study the surface and edge properties, a tight-binding model was derived from our \textit{ab initio } calculations using W $d$ and S $p$ orbitals by computing the maximally localized Wannier functions (MLWFs) using the {\sc wannier90} code~\cite{mostofi2014updated}.
To obtain the surface spectrum, the method of calculating the surface Green's function using iterative Green's functions~\cite{PhysRevB.28.4397,Sancho_1984,Sancho_1985} was used, as implemented in the WannierTools code~\cite{WU2017}.   
Further analysis of these surface states and their underlying topological properties was facilitated by the WannierTools code~\cite{WU2017}.  \\

To characterize the topological nature of our investigated systems, we used the approach proposed by Rui \textit{et al.}~\cite{Z2_Rui} and Soluyanov \textit{et al.}~\cite{soluyanov2011computing}, which is a computationally feasible method to calculate the $Z_2$ topological invariant, related to the Wannier charge centers (WCCs). We very briefly summarize their method.
In any unit cell $R$, Wannier functions (WFs) are given by

\begin{equation}
    \big|R_n\big \rangle = \frac{1}{2\pi} \int_{-\pi}^{\pi} dk~e^{-ik(R-x)}~\big|u_{nk}\big \rangle,
\end{equation}

where $\big|u_{nk}\big \rangle$ is the periodic part of the Bloch state of band $n$ at momentum $k$. 
WCCs are obtained by taking an expectation value of the position operator in the state $\big| 0_n \big\rangle$ corresponding to WFs in the unit cell $R = 0$. A tight binding Hamiltonian is derived based on our density functional results, and is used to generate MLWFs using {\sc wannier90} code~\cite{mostofi2014updated}.
The WCCs of these MLWFs are obtained, and the evolution of these WCCs along a fixed momentum plane is tracked to check for the even or odd value of the $Z_2$ topological invariant. \\

Even under time-reversal symmetry (TRS), non-centrosymmetric systems were theorised to show a Hall effect in the non-linear regime due to the presence of a Berry Curvature Dipole (BCD).
BCD is a measure of the first order moment of the Berry curvature and is given by~\cite{PhysRevLett.115.216806}

\begin{equation}
    D_{ab} = \int_{k}~f_n^0(\textbf{k})~\frac{\partial\Omega^n_b}{\partial k_a}.
\end{equation}

Here, $f_n^0$(\textbf{k}) is the equilibrium Fermi-Dirac distribution and $\Omega_b^n$ the Berry curvature. 
In a two-dimensional system, $\Omega^n$ only has an out-of-plane component, and is given by

\begin{equation}
    \Omega_x^n (\textbf{k}) =  2i\hbar^2 \sum_{m \neq n} \frac{\langle n|\Hat{v}_y|m\rangle \langle n|\Hat{v}_z|m\rangle}{(\epsilon_n - \epsilon_m)^2},
\end{equation}

where $\epsilon_m$ and $|m\rangle$ are the energy eigenvalue and eigenvector of the Hamiltonian and $\hat{v}_\alpha$ is the velocity operator along $\alpha$. Here $\hbar$ is the reduced Planck's constant.
We used the wannier-based tight-binding Hamiltonian to compute the Berry curvature and BCD associated with it using the {\sc wannier-berri} code~\cite{tsirkin2020high}.

\section{\label{sec:bulk}Bulk WS$_2$}


\subsection{\label{subsec:struct} Structural Properties}

\begin{figure}
\begin{center}
  \includegraphics[scale=0.6]{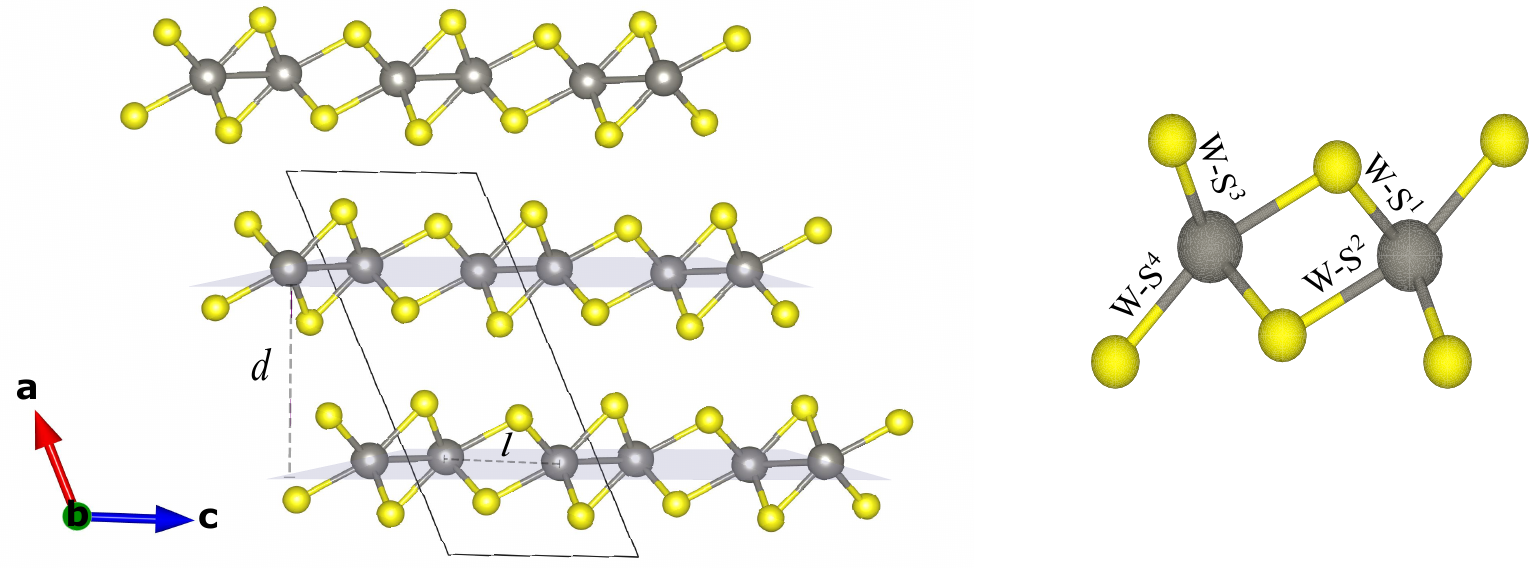}
   \caption{Structure of the 2M phase of WS$_2$. The inter-layer distance is marked $d$ and W-W zigzag chain separation is marked $l$. The different types of W-S bonds are also marked in the zoom of the structure in the right panel. Here the grey spheres depict the W atoms, while the yellow spheres represent the S atoms.} \label{structure}
\end{center}
\end{figure}

In this section, we begin with a detailed analysis of the structural properties of the newly discovered 2M phase of WS$_2$. We compare and contrast the results from different exchange-correlation functionals that were used.
2M WS$_2$ was experimentally determined to form in the monoclinic $C_2/m$ space group \cite{fang2019discovery}. This phase is similar to the previously known 1T' and T$_d$ phases of other TMDCs, since the basic units are distorted octahedra of W and S, and yet it is different in terms of the stacking along the long axis $a$.
The stacking is such that the adjacent WS$_2$ monolayers are displaced by a distance of $a/2$, as illustrated in Fig.~\ref{structure}. 
The primitive unit cell consist of 6 atoms with 4 different types of W-S bonds. We label these as W-S$^1$, W-S$^2$, W-S$^3$, W-S$^4$, as shown in Fig.~\ref{structure}. The distorted 1T' structure has a characteristic in-plane W-W zigzag chain. We denote the zigzag chain separation by $l$, while the interlayer separation is labeled by $d$. \\

Starting from the experimentally-reported geometry, we performed three separate relaxation calculations using LDA, GGA-PBE and GGA-PBE with Grimme's DFT-D2 correction.
We extracted the different structural parameters for all the three relaxed structures and compared them to the crystal structure obtained experimentally. 
In Table~\ref{tab:layer} we present the inter-layer distance ($d$), W-W bond length and W zigzag chain separation ($l$) obtained for the four structures. 
The inter-layer distance was calculated as perpendicular distance between the two planes occupied by W layers (see Fig.~\ref{structure}). \\

\begin{table}
  \centering
    \caption{Comparison of structural properties - inter-layer distance ($d$), W-W bond length and W zigzag chain separation ($l$) - of bulk 2M WS$_2$ obtained from experiment and from first-principles computations employing different exchange correlation functionals: local density approximation (LDA), generalized gradient approximation (GGA-PBE) and generalized gradient approximation with Grimme's DFT-D2 correction. All distances are in Angstrom. Experimental structural data is taken from Reference \onlinecite{fang2019discovery}.}
    \label{tab:layer}
    \begin{ruledtabular}
    \begin{tabular}{lccc}
      & \textbf{\textit{d}} & \textbf{W-W bond length} & \textbf{\textit{l}}  \\ 
        \hline \\
      Experiment \cite{fang2019discovery} & 6.0580 & 2.7940 & 3.7940 \\
       &  & & \\
      LDA & 6.0611 & 2.7764 & 3.8153 \\
      GGA-PBE & 6.0674 & 2.7919 & 3.7990   \\
      GGA-PBE (DFT-D2) & 6.0644 & 2.8078 & 3.7813  \\
    \end{tabular}
    \end{ruledtabular}
 \end{table}
 
Comparing the three cases, from Table~\ref{tab:layer}, it can be noted that all three calculations give similar inter-layer distances $l$, with an error of less than 0.2\% compared to the experimentally reported value. 
In case of the different bond lengths obtained, LDA over-binds and GGA under-binds the atoms when compared to experimentally observed values, as is traditionally expected \cite{martin_2004}. 
From Table~\ref{tab:bonds}, it can be inferred that GGA-PBE gives a marginally closer estimate of experimental bond length than LDA does. 
All W-S bond lengths obtained using GGA-PBE functional (both with and without Grimme's DFT-D2 correction) are less than 0.5\% off of the experimentally determined bond lengths. 
This is also true for the case of $d$ and W-W separation, presented in Table~\ref{tab:layer}. \\

\begin{table}
  \centering
    \caption{Comparison of different W-S bond lengths in bulk 2M WS$_2$ obtained from experiment and from first-principles computations employing different exchange correlation functionals: local density approximation (LDA), generalized gradient approximation (GGA-PBE) and generalized gradient approximation with Grimme's DFT-D2 correction. All distances are in Angstrom. Experimental structural data is obtained from Reference \onlinecite{fang2019discovery}.}
    \label{tab:bonds}
    \begin{ruledtabular}
    \begin{tabular}{l c c c c} 
     & \multicolumn{4}{c}{\textbf{Bond Lengths}}  \\ 
       & \textbf{W-S$^1$} & \textbf{W-S$^2$} & \textbf{W-S$^3$} & \textbf{W-S$^4$}\\
        \hline \\
      Experiment\cite{fang2019discovery} & 2.4570 & 2.5150 & 2.3960 & 2.4150 \\
       & & & &\\
      LDA & 2.4418 & 2.5056 & 2.3756 & 2.3968\\
      GGA-PBE & 2.4637 & 2.5118 & 2.3982 & 2.4205  \\
      GGA-PBE (DFT-D2) & 2.4617 & 2.5080 & 2.3988 & 2.4183 \\
    \end{tabular}
    \end{ruledtabular}
 \end{table}


\subsection{\label{subsec:electronic} Electronic Properties}

\begin{figure}
\begin{center}
  \includegraphics[scale=0.6]{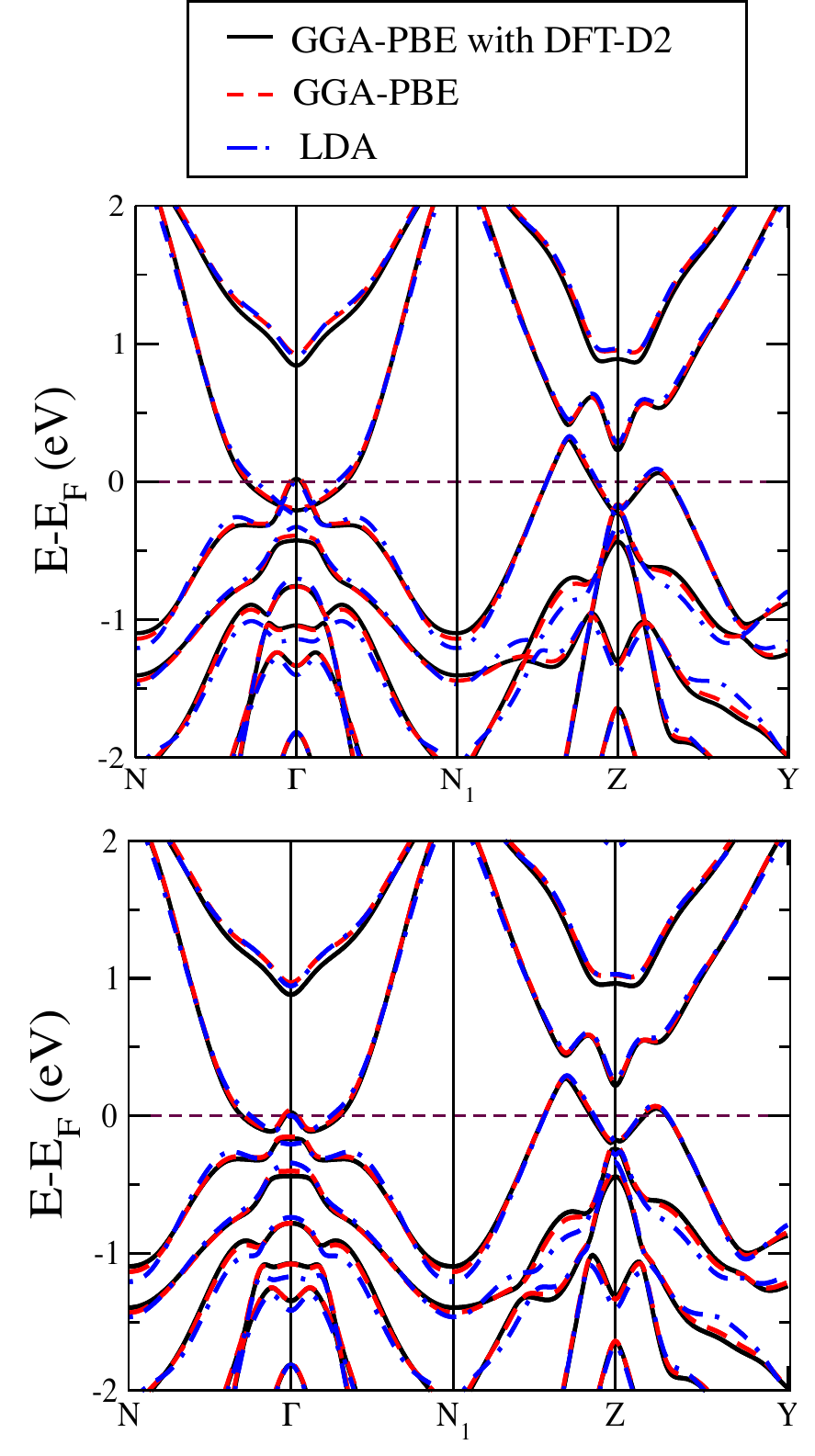}
   \caption{A comparison of band structures obtained without (above) and with (below) spin-orbit coupling effects employing different exchange correlation functionals: local density approximation (LDA), generalized gradient approximation (GGA-PBE) and generalized gradient approximation with Grimme's DFT-D2 correction. The system is semi-metallic both without and with spin-orbit coupling.}\label{band-compare}  
\end{center}
\end{figure}

After an analysis of the structural properties of bulk 2M WS$_2$, we next move on to the study of its electronic properties.
We begin by examining the band structure with and without spin-orbit coupling.
A comparison of the band structures obtained using the three different exchange correlation functionals is shown in Fig.~\ref{band-compare}. 
We find that different choices of the exchange-correlation functional leave the band structure around the Fermi level nearly unchanged.
For all three functionals, we find that the system is a semi-metal, with a narrow electron pocket around $\Gamma$ and a wider hole pocket along N$_1$-Z. 
A smaller hole pocket is also present along the Z-Y direction.
Including spin-orbit coupling effects retains the electron and hole pockets, but the fundamental bandgap increases with a 0.05-0.08 eV bandgap induced away from $\Gamma$. 
Including SOC effects also split the band crossings around -0.2 eV below Fermi level at Z. 
All the three functionals (LDA, GGA-PBE and GGA-PBE including Grimme's DFT-D2 van der Waals correction), capture these effects around the Fermi level in a similar way, but the fundamental bandgap  and inverted bandgap at $\Gamma$  decrease as one goes from LDA to GGA-PBE as summarized in Table~\ref{tab:bandgap}.
We note that our calculated value of inverted bandgap is similar to previous calculations~\cite{Qian1344} for 1T' WS$_2$ which places this value to be around 0.2 eV. 
This is expected since the bulk 2M WS$_2$ is constituted of 1T' monolayers. 
Including SOC effects increases the bandgap of the system, as mentioned above, and reduces the inverted bandgap at $\Gamma$.   \\

\begin{table*}
  \centering
    \caption{The fundamental bandgap (with SOC) and inverted bandgap at $\Gamma$ calculated using different exchange correlation functionals - local density approximation (LDA) and generalized gradient approximation (GGA-PBE) - and also generalized gradient approximation with Grimme's DFT-D2 correction. All energies are in eV.}
    \label{tab:bandgap}
    \begin{ruledtabular}
    \begin{tabular}{l c c c}
      &  \textbf{Fundamental gap} & \multicolumn{2}{c}{\textbf{Inverted gap at $\Gamma$}}  \\
       &  &  \textbf{without SOC} & \textbf{with SOC} \\
         \hline
       \\
      LDA & 0.082 & 0.234 & 0.208 \\
      GGA-PBE & 0.050 & 0.230 &0.195 \\
      GGA-PBE (DFT-D2) & 0.055 & 0.229 & 0.190 \\
    \end{tabular}
    \end{ruledtabular}
 \end{table*} 

Given the reasonable performance of the GGA-PBE functional and considering that the layers are held together by weak van der Waals forces, semiempirical Grimme's DFT-D2 corrections were included for all following calculations of bulk and bilayer WS$_2$ along with GGA-PBE exchange-correlation functional, including SOC effects. \\

\begin{figure*}
\begin{center}
  \includegraphics[scale=0.7]{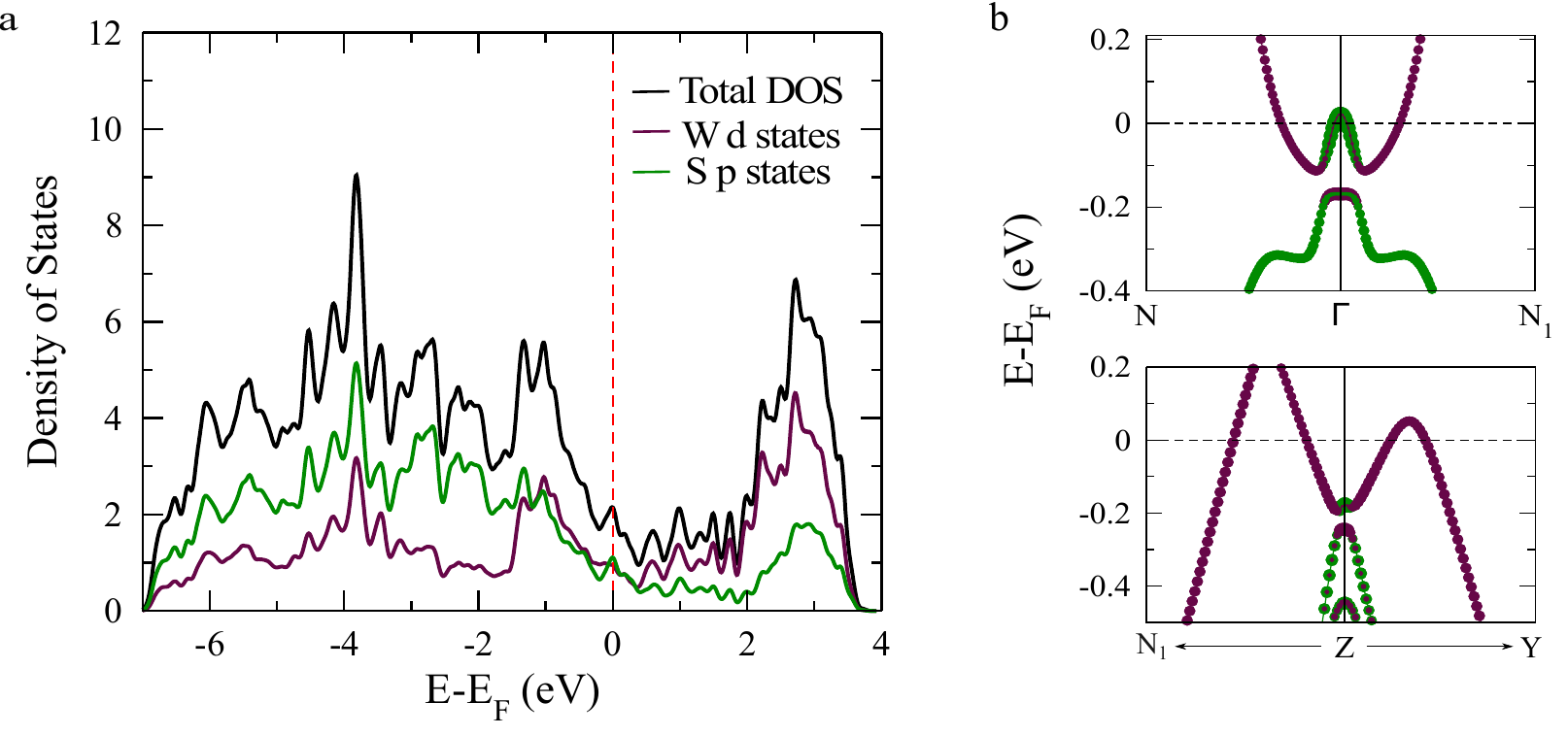}
   \caption{(a) Total and projected density of states for bulk 2M WS$_2$. Around the Fermi level, nearly all contributions are from the W $d$ and S $p$ states. (b) W $d$ and S $p$ states projected onto electronic bands, zoomed around $\Gamma$ and Z points. Note the band inversions occurring around both these points.}\label{bulkdos} 
\end{center}
\end{figure*}

The inverted band features around $\Gamma$ and Z are suggestive of possible topological nature of this material.
The density of states (DOS) around Fermi level [Fig.~\ref{bulkdos} (a)] confirms the (semi)metallic nature of the material with the states having a larger S $p$ character (green) below Fermi level and W $d$ character (maroon) above Fermi level. 
Overall, most part of the valence states and low energy conduction states are dominated by W $d$ states and S $p$ states. 
Keeping this in mind, the presence of band inversion can be investigated by projecting these underlying atomic-like orbitals onto the bands. 
As shown in Fig.~\ref{bulkdos} (b), around the Fermi level, the W $d$ states (in maroon) and S $p$ states (in green) are inverted both at $\Gamma$ and Z.
The band inversions at $\Gamma$ is different from the one at Z in the sense that these occur between two different pairs of band. 
Hence, this is an indication that the system is a strong topological material, rather than a weak one, which was further confirmed from the topological invariant calculation, as we will discuss next. \\

\begin{figure}
\begin{center}
  \includegraphics[scale=0.5]{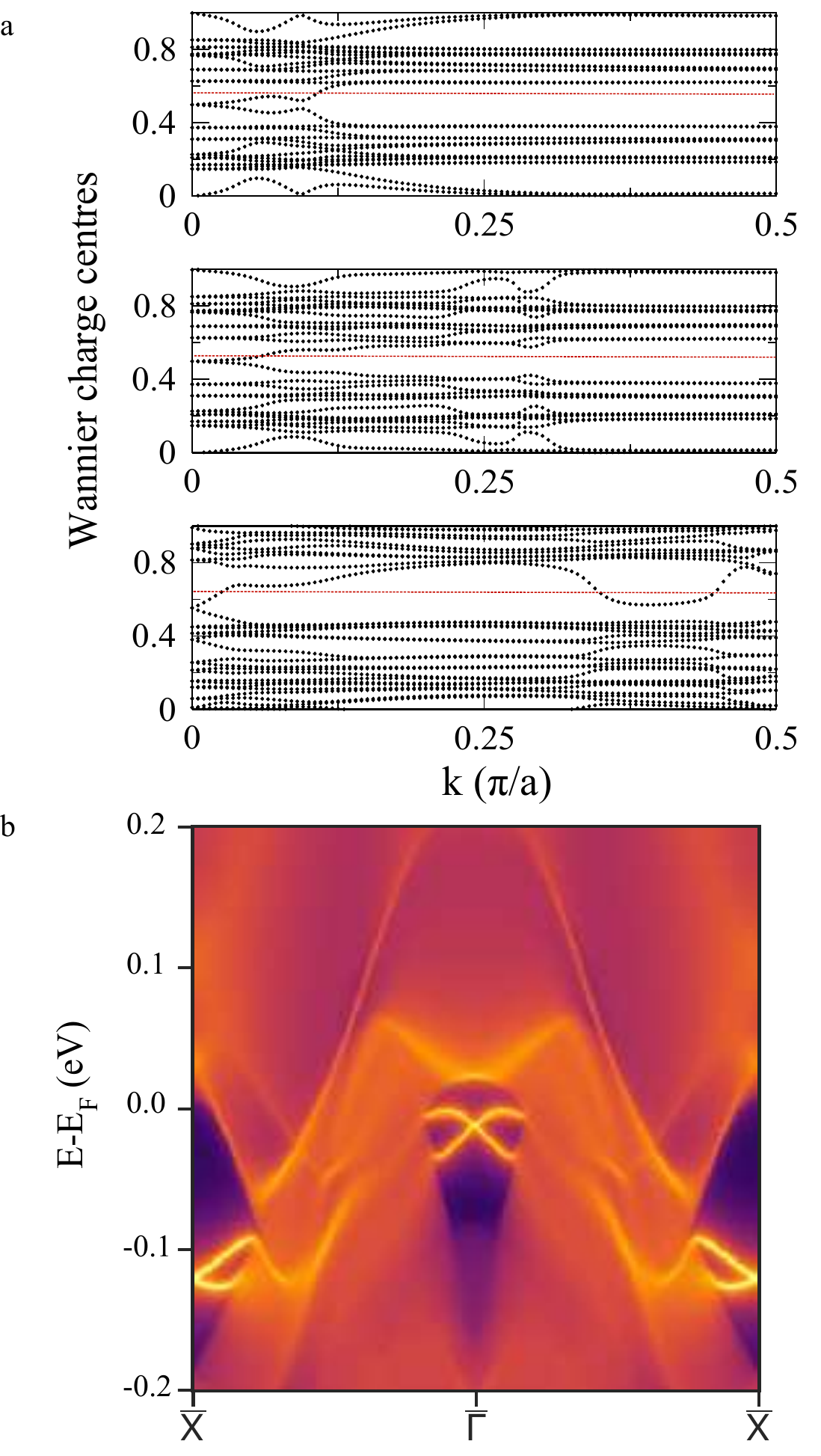}
   \caption{Topological nature of bulk 2M WS$_2$. (a) Wannier charge centres plotted along (top to bottom) $k_x=0$ , $k_y=0$, $k_z=0$ planes. Any line (red) drawn parallel to $k$ axis intersects the wannier charge centres an odd number of times revealing an odd principal Z$_2$ invariant. (b) Surface states on the (100) surface. The system has a strong topological character with a surface Dirac cone at $\Gamma$.} \label{bulkSS}
\end{center}
\end{figure}

To characterize the topological nature of bulk 2M WS$_2$ we use the WCCs, employing the methodology described in Section~\ref{sec:methods}. The WCCs for different momentum planes are shown in Figure~\ref{bulkSS} (a). 
Each panel in Figure~\ref{bulkSS} (a) refers to each of the three different time reversal invariant planes -- $k_x = 0$, $k_y = 0$ and $k_z = 0$ -- along which we track the evolution of WCCs.
The line (red dashed) drawn parallel to $k$ axis intersects the WCCs (black dotted) an odd number of times in all the $k_x = 0$, $k_y = 0$ and $k_z = 0$ planes. 
Our overall calculated topological invariant is (1;000), which confirms that the system is indeed a strong topological material.
The odd principal topological invariant also means that there exist topologically protected states on the surface of the material. 
These are another important signature of the non-trivial topological nature of a material.
We used MLWFs derived from the density functional calculations to compute the surface spectrum on the (100) surface, using a recursive Green's functions method (see Section~\ref{sec:methods}). The result is plotted in Figure~\ref{bulkSS} (b). Indeed, we find that there exists a surface state close to the Fermi level. Around $\Gamma$ it takes the form of a Dirac cone, which merges into bulk states as one moves away from the Brillouin zone center. Hence, consistent with previous computations~\cite{fang2019discovery}, we confirm that the new monoclinic phase of WS$_2$ should harbor topologically non-trivial surface states.
Angle resolved photoemission experiments will be able to directly probe these and test our predictions. \\

\section{\label{sec:bilayer}Bilayer 2M WS$_2$}
 
Upon the discovery of the new 2M phase of bulk WS$_2$, several natural questions arise: What are the properties of few layer 2M phases? Are they topological in nature? In this section we explore these questions focusing on the properties of bilayer 2M WS$_2$. 
The unit cell of the 2M structure comprises of monolayers shifted by $a/2$ along the stacking direction. So, the smallest few layer unit we consider is the bilayer. 


\subsection{Structural Properties}

\begin{figure}
\begin{center}
  \includegraphics[scale=0.4]{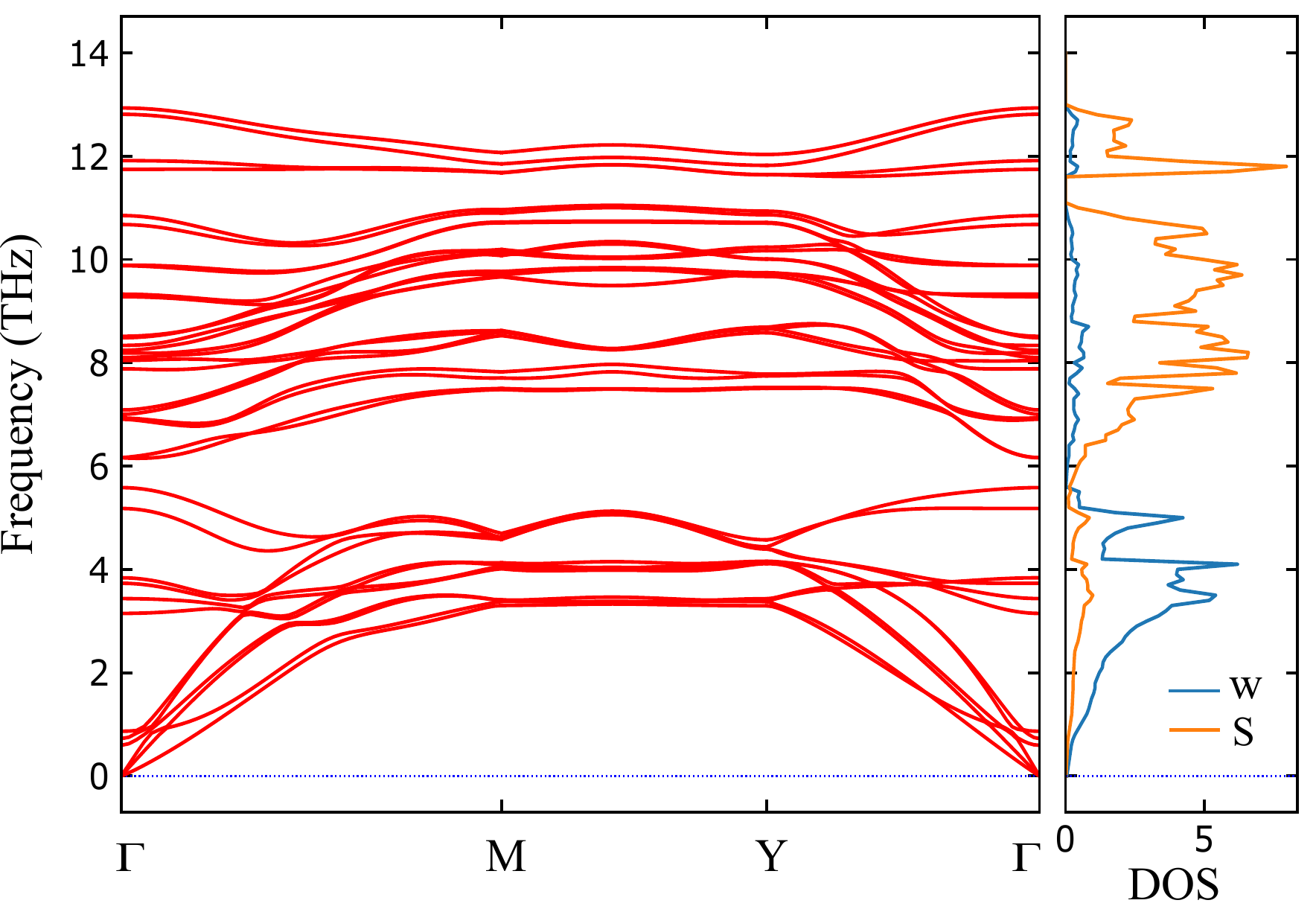}
   \caption{Phonon dispersion and projected phonon density of states for bilayer 2M WS$_2$. Absence of any imaginary frequencies indicates that the bilayer structure is dynamically stable. The projected phonon density of states reveals a larger contribution of W displacements at lower phonon frequencies, while S atoms contribute more at higher frequencies.}\label{phonon_bands} 
\end{center}
\end{figure}

The first question that then arises is whether the bilayer is structurally stable or not. 
To answer this question, we carried out phonon calculations using the frozen-phonon approach as described in Section~\ref{sec:methods}). 
Our calculated phonon bands are presented in Fig.~\ref{phonon_bands}. Most notably, we find that the phonon frequencies are real across the phonon momenta. 
This means that the bilayer of 2M WS$_2$ is dynamically stable. 
The analysis of the projected phonon density of states (see Fig.~\ref{phonon_bands}, right panel) shows that the displacement of the heavier W atoms contributes strongly to the phonon modes up to frequencies of around 6 THz. 
Contribution from the lighter S atoms is larger from 6 THz to the top of the dispersion range ($\sim$13 THz). The zone centre ($\Gamma$) optical phonon frequencies and the associated irreducible representations (IRRs) are listed in Table~\ref{tab:phonon_freq}. 
The 33 optical phonon modes are classified into 22 A' and 11 A'' modes, with all of those listed being both Raman and infrared active.
These signatures can be helpful in facilitating experimental identification of the bilayer 2M WS$_2$ structure. 

\begin{table}
  \centering
    \caption{Optical phonon frequencies ($\nu$) at $\Gamma$ point for bilayer 2M WS$_2$. The modes are classified into 22 A' and 11 A'' symmetry modes, with all 33 modes being both infrared and Raman active. Here IRR stands for irreducible representation.}
    \label{tab:phonon_freq}
    \begin{ruledtabular}
    \begin{tabular}{c c c c c c}
        \textbf{No.} & \textbf{IRR} & \textbf{$\nu$ (THz)} & \textbf{No.} & \textbf{IRR} & \textbf{$\nu$ (THz)} \\
         \hline
         
        1   &  A'    &        0.595    &   18  &  A''   &        8.099  \\ 
        2   &  A''   &        0.724    &   19  &  A''   &        8.187  \\
        3   &  A'    &        0.867    &   20  &  A'    &        8.241  \\
        4   &  A''   &        3.148    &   21  &  A'    &        8.336  \\
        5   &  A''   &        3.435    &   22  &  A''   &        8.484  \\
        6   &  A'    &        3.731    &   23  &  A''   &        8.512  \\
        7   &  A'    &        3.838    &   24  &  A'    &        9.278  \\
        8   &  A'    &        5.180    &   25  &  A'    &        9.328  \\
        9   &  A'    &        5.584    &   26  &  A'    &        9.880  \\
        10  &  A''   &        6.161    &   27  &  A'    &        9.888  \\
        11  &  A''   &        6.163    &   28  &  A'    &        10.676 \\
        12  &  A'    &        6.907    &   29  &  A'    &        10.850 \\
        13  &  A'    &        6.930    &   30  &  A'    &        11.744 \\
        14  &  A''   &        7.000    &   31  &  A'    &        11.913 \\
        15  &  A''   &        7.085    &   32  &  A'    &        12.810 \\
        16  &  A'    &        7.883    &   33  &  A'    &        12.933 \\ 
        17  &  A'    &        8.044    & & &

    \end{tabular}
    \end{ruledtabular}
\end{table} 
 

 \subsection{Electronic Properties}
 
\begin{figure}
\begin{center}
  \includegraphics[scale=0.5]{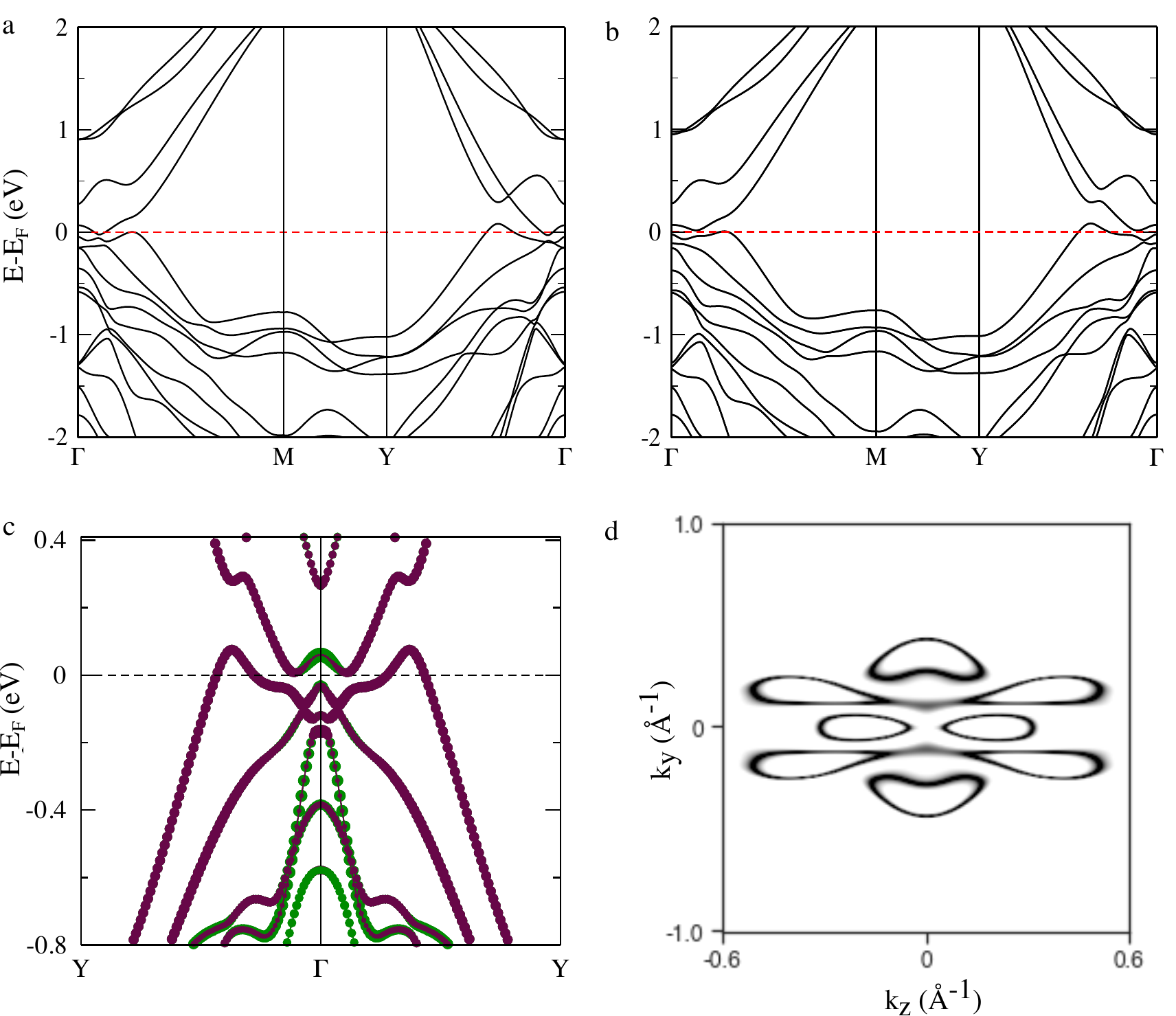}
   \caption{Electronic band structure of bilayer 2M WS$_2$. Similar to the bulk, bilayer also shows a semi-metallic character. The band structure (a) without and (b) with spin orbit coupling. (c) The bands projected onto W $d$ states (maroon) and S $p$ states (green) clearly show band inversion at $\Gamma$. (d) The Fermi surface plot in the $k_x=0$ plane showing the electron and hole pockets. The double dumbbell-like feature arises due to the electron pocket, while the dumbbell-like feature along with the cap structures above and below the electron pocket come from the hole pocket.} \label{bilayer_bands}
\end{center}
\end{figure}

Having established the structural stability of the 2M WS$_2$ bilayer, we next investigate its electronic properties.
The electronic band structure and Fermi surface reveals the bilayer to be similar to the bulk with a semi-metallic character. The bilayer displays a zero indirect bandgap but finite direct gap at every $k$-point (see Figure~\ref{bilayer_bands}).
A comparison between the band structures with and without SOC effects reveals that several band degeneracies are lifted in both the valence and conduction bands [see Figure~\ref{bilayer_bands} (a) and (b)]. An electron pocket around $\Gamma$ vanishes upon inclusion of SOC, while the bandgap of $\sim$ 0.1 eV remains almost unchanged at the Brillouin zone center. 
To ascertain the orbital nature of bands around the Fermi level and look for possible band inversions in the bilayer system, we carried out a projected bands analysis.
Understanding that, similar to the bulk, both W $d$ states and S $p$ states contribute most around the Fermi level, the projection of these states onto the electronic band structure is a straight-forward method to identify band inversions.
The projected bands along the Y-$\Gamma$-Y direction, as shown in Figure~\ref{bilayer_bands} (c), reveal a clear inversion between W (maroon) and S (green) bands around $\Gamma$.
The inversion of states, even though small, is similar to the ones present in the bulk, hence leading to the suggestion that even in lower dimensions, the 2M phase has topologically non-trivial edge states. \\

\begin{figure*}
\begin{center}
  \includegraphics[scale=0.46]{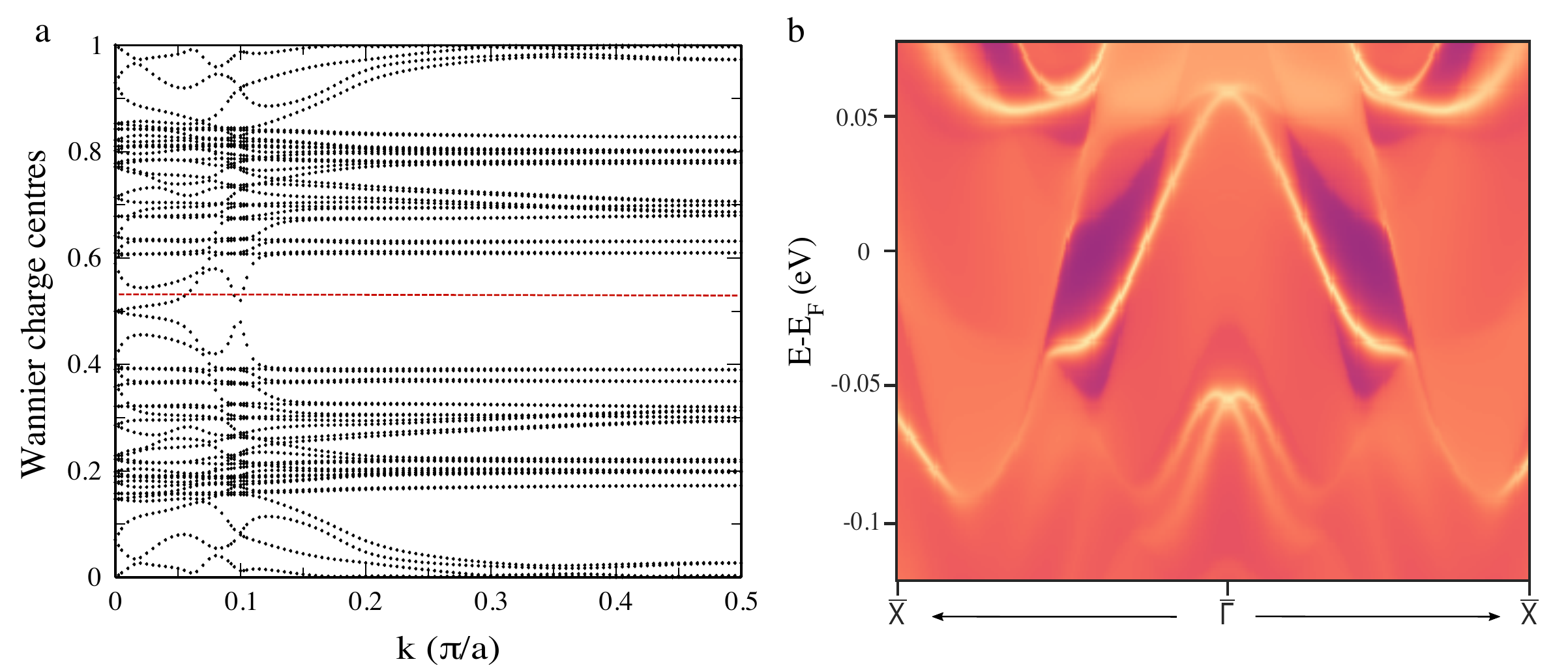}
   \caption{Non-trivial topology in bilayer 2M WS$_2$. (a) WCCs plotted for $k_x = 0$ plane. A line (red) drawn parallel to $k$ axis intersects the WCCs an odd number of times. Hence, the material is topologically non-trivial. (b) Calculated edge states in the system. The brighter bands indicate the higher density of states along the edges of the material. Note that the edge states are mixed heavily with bulk states.} \label{bilayerSS}
\end{center}
\end{figure*}

To further understand the topological nature of bilayer 2M WS$_2$, we calculated the Z$_2$ topological invariant for our system.
Similar to the bulk, we constructed a tight-binding model based on Wannier functions employing the W $d$ and S $p$ states.  
Using this tight-binding model, we tracked the WCCs along a fixed $k_x=0$ plane [Figure~\ref{bilayerSS} (a)]. 
As shown in Figure~\ref{bilayerSS} (a), we discover that the line (red dashed) drawn parallel to $k$ axis intersects the WCCs (black dotted) an odd number of times.
Thereby we obtain that the bilayer 2M WS$_2$ has an odd $Z_2$ number ($\nu_0 =1$) and is therefore topologically non-trivial.
To characterize the edge spectrum of the bilayer, we consider a semi infinite one-dimensional ribbon geometry and compute the edge spectrum using the recursive Green's function technique.
Figure~\ref{bilayerSS} (b) shows the edge sates associated with the bilayer.
We find clear signatures of the edge states close to the Fermi level.
The edge states around 0.05 eV above the Fermi level lie close to the Brillouin zone center. Remarkably, it inherits a shape similar to the surface Dirac cone observed in the bulk phase.
Interestingly, we find another edge band crossing around 0.05 eV below the Fermi level can be attributed to a Rashba-like splitting of bands, due to the broken inversion symmetry in the bilayer.
Therefore, based on first-principles calculations, we have successfully predicted a new two-dimensional topological material in the TMDC family. 


\begin{figure*}[p]
\begin{center}
  \includegraphics[scale=0.8]{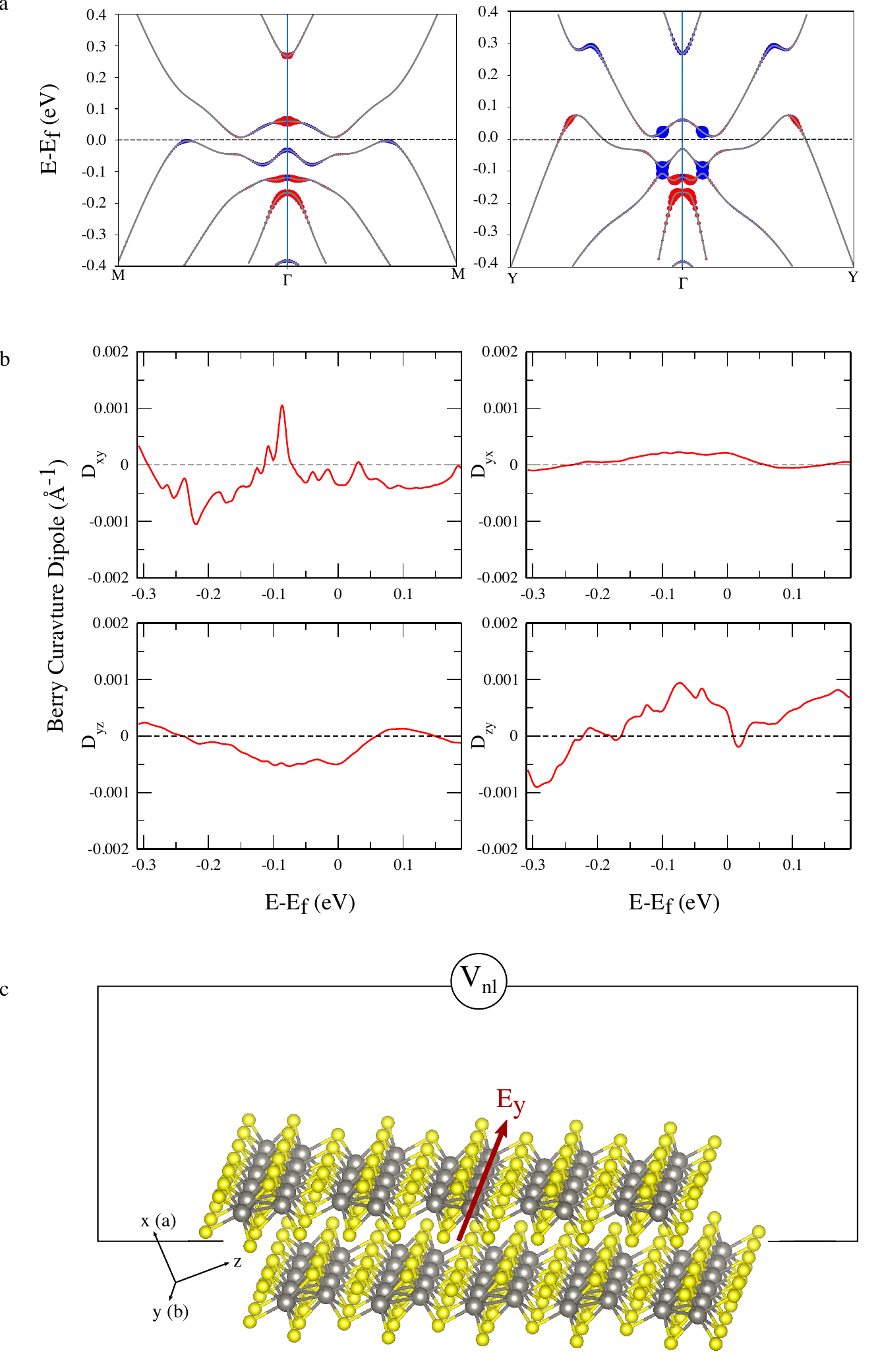}
   \caption{Berry curvature and its dipoles. (a) The Berry curvature of bilayer 2M WS$_2$ along M-$\Gamma$-M and Y-$\Gamma$-Y directions. Here red and blue colors indicate positive and negative values respectively. (b) The BCD associated with bilayer 2M WS$_2$. The BCD tensor, in this case, has four non-zero components - $D_{xy}$, $D_{yx}$, $D_{yz}$ and $D_{zy}$. The dotted line marks the BCD associated with bulk. (c) NLAHE in bilayer 2M WS$_2$. When an electric field, $E_y$, is applied along the $y$ axis, a non-linear Hall-like voltage $V_{nl}$ is generated perpendicular to it.} \label{bcd_bilayer}
\end{center}
\end{figure*}

\subsection{Berry Curvature Dipole in bilayer 2M WS$_2$}

Recent theories have suggested that materials having reduced symmetry display Hall effect in the non-linear regime, even while TRS remains preserved~\cite{PhysRevLett.115.216806}.
Such a non-linear response was attributed to a dipole moment in the momentum space associated with the Berry curvature. 
In systems that break inversion symmetry, a dipole moment can arise due to separation of positive and negative Berry curvatures along different momentum directions.
This Berry Curvature Dipole (BCD) leads to several non-linear response phenomena, whose signatures have been recently measured experimentally~\cite{xu2018electrically} \\

The bulk structure of 2M WS$_2$, with its monoclinic $C_{2/m}$ space group, preserves inversion symmetry.
The coexistence of inversion symmetry and TRS leads the Berry curvature and hence, the associated BCD, to be vanish.
The transition from bulk to bilayer breaks this inversion symmetry, and the space group is reduced to $C_m$.
The weakly broken inversion symmetry is reflected in the band structure as degeneracy of bands being lightly lifted along $\Gamma-Y$ and $\Gamma-M$.
This non-centrosymmetric structure leads to the formation of BCD in the bilayer and sets it apart from the parent bulk 2M WS$_2$.
The only symmetry element present in the bilayer is a mirror plane $M_y : y \xrightarrow{}-y$.
The symmetry is important in determining the non-zero components of the BCD tensor and the resulting non-linear anomalous Hall effect (NLAHE)~\cite{PhysRevLett.125.046402,zhang2018electrically,ma2019observation}. \\

We start by examining the distribution of the Berry curvature in the non-centrosymmetric bilayer. Fig~\ref{bcd_bilayer} (a) shows the distribution of Berry curvature along $Y-\Gamma-Y$ and $M-\Gamma-M$ directions (red and blue colors indicate the positive and negative values respectively).
The Berry curvature is enhanced and mostly concentrated around the regions where the bands almost cross.
The monoclinic crystal lattice of the bilayer is such that the $x$ and $y$ cartesian axes align with the $a$ and $b$ lattice vectors (Fig.~\ref{bcd_bilayer}) and the $c$ vector has components along both $x$ and $z$ direction.
As a result of this structure and due to the mirror plane $M_y$ present in the bilayer, all components of BCD except $D_{xy}$, $D_{yx}$, $D_{yz}$ and $D_{zy}$ vanish.
Fig~\ref{bcd_bilayer}(b) shows the four non-zero components of the BCD tensor as a function of the Fermi level.
The $D_{xy}$ displays oscillating behaviour with energy, while other components have a slighlty more smoother variation. 
The figure shows that BCD dips and is negative near Fermi level for the $D_{xy}$ and $D_{yz}$ components, while it slightly peaks and is positive for $D_{yx}$ and $D_{zy}$ components.
The estimated values of BCD at Fermi level are $-3.52\times10^{-4}$, $2.10\times10^{-4}$, $-4.97\times10^{-4}$ and $4.15\times10^{-4}$ for $D_{xy}$, $D_{yx}$, $D_{yz}$ and $D_{zy}$ respectively. \\

BCDs are interesting since they are associated with several non-linear responses such as non-linear Hall currents, second harmonic generation and photocurrents from circular photo-galvanic effect~\cite{PhysRevLett.115.216806,ma2017direct,sun2017circular,ma2019observation}.
An application of electric field along the $y$ axis, parallel to the generated BCD, i.e., along the direction of W-W zig-zag chain [Fig.~\ref{bcd_bilayer}(c)], leads to a non-linear Hall-like voltage $V_{nl}$ generated perpendicular to the field, i.e., along $c$.
The NLAHE exist without breaking the time-reversal symmetry and has been previously predicted in similar layered systems such as monolayer WTe$_2$ and MoTe$_2$~\cite{zhang2018electrically}, resulting from the formation of BCD due to breaking of inversion symmetry.
Further, we find that this NLAHE signal obtained is strongly anisotropic.
We hope that these non-linear responses and BCD can be measured experimentally for our system, as was recently done for bilayer WTe$_2$~\cite{ma2019observation}. \\

We mention in passing that it would be interesting to determine experimentally, whether such low-dimensional phases of 2M WS$_2$ also exhibit superconductivity and whether they retain the relatively high superconducting critical temperature of their parent bulk phase. If found to be superconducting, they could become an intriguing playground to study the interplay of topology, superconductivity and proximity effects in reduced dimensions~\cite{xu2014artificial,kim2017strong,huang2018inducing,huang2020edge,trainer2020proximity,lupke2020proximity}. \\

\section{\label{sec:conclusion}Summary}

In summary, we investigated the properties of bulk and bilayer 2M WS$_2$ using first-principles density functional calculations. 
We systematically investigated the structural properties of the bulk compound, obtained using different types of exchange correlation functionals and their comparison to the experimentally reported values. 
We presented the electronic properties of the bulk phase highlighting the effects of spin-orbit coupling. 
We analyzed the topological properties of the bulk phase, and demonstrated its non-trivial topological nature. 
We predicted bilayer 2M WS$_2$ as a new two-dimensional TMDC, which features topological band inversions. 
We demonstrated the dynamical stability of this new structure and presented the computed phonon modes and their symmetries, which may assist in experimental identification of this phase. 
We investigated the electronic properties of this new system, highlighting the band inversions between W $d$ states and S $p$ states near the Brillouin zone center. 
We showed that bilayer 2M WS$_2$ exhibits a non-trivial topology, by means of topological invariant computations and explicit calculation of the edge states. 
We also demonstrate that the Berry curvature dipole in the bilayer system is non-zero due to the broken inversion symmetry, in contrast to the bulk. We calculated its variation with the Fermi energy and propose a setup to measure it.
We hope that our predictions lead to the experimental realization of this new two-dimensional topological material.\\

\section*{Acknowldegments}

We thank Nagaphani Aetukuri and Manish Jain for useful comments. N.B.J. acknowledges support from the Prime Minister's Research Fellowship. A.N. acknowledges support from the start-up grant (SG/MHRD-19-0001) of the Indian Institute of Science and DST-SERB (project number SRG/2020/000153).


\begin{thebibliography}{56}%
\makeatletter
\providecommand \@ifxundefined [1]{%
 \@ifx{#1\undefined}
}%
\providecommand \@ifnum [1]{%
 \ifnum #1\expandafter \@firstoftwo
 \else \expandafter \@secondoftwo
 \fi
}%
\providecommand \@ifx [1]{%
 \ifx #1\expandafter \@firstoftwo
 \else \expandafter \@secondoftwo
 \fi
}%
\providecommand \natexlab [1]{#1}%
\providecommand \enquote  [1]{``#1''}%
\providecommand \bibnamefont  [1]{#1}%
\providecommand \bibfnamefont [1]{#1}%
\providecommand \citenamefont [1]{#1}%
\providecommand \href@noop [0]{\@secondoftwo}%
\providecommand \href [0]{\begingroup \@sanitize@url \@href}%
\providecommand \@href[1]{\@@startlink{#1}\@@href}%
\providecommand \@@href[1]{\endgroup#1\@@endlink}%
\providecommand \@sanitize@url [0]{\catcode `\\12\catcode `\$12\catcode
  `\&12\catcode `\#12\catcode `\^12\catcode `\_12\catcode `\%12\relax}%
\providecommand \@@startlink[1]{}%
\providecommand \@@endlink[0]{}%
\providecommand \url  [0]{\begingroup\@sanitize@url \@url }%
\providecommand \@url [1]{\endgroup\@href {#1}{\urlprefix }}%
\providecommand \urlprefix  [0]{URL }%
\providecommand \Eprint [0]{\href }%
\providecommand \doibase [0]{http://dx.doi.org/}%
\providecommand \selectlanguage [0]{\@gobble}%
\providecommand \bibinfo  [0]{\@secondoftwo}%
\providecommand \bibfield  [0]{\@secondoftwo}%
\providecommand \translation [1]{[#1]}%
\providecommand \BibitemOpen [0]{}%
\providecommand \bibitemStop [0]{}%
\providecommand \bibitemNoStop [0]{.\EOS\space}%
\providecommand \EOS [0]{\spacefactor3000\relax}%
\providecommand \BibitemShut  [1]{\csname bibitem#1\endcsname}%
\let\auto@bib@innerbib\@empty
\bibitem [{\citenamefont {Manzeli}\ \emph {et~al.}(2017)\citenamefont
  {Manzeli}, \citenamefont {Ovchinnikov}, \citenamefont {Pasquier},
  \citenamefont {Yazyev},\ and\ \citenamefont {Kis}}]{2D_TMD_natrev2017}%
  \BibitemOpen
  \bibfield  {author} {\bibinfo {author} {\bibfnamefont {S.}~\bibnamefont
  {Manzeli}}, \bibinfo {author} {\bibfnamefont {D.}~\bibnamefont
  {Ovchinnikov}}, \bibinfo {author} {\bibfnamefont {D.}~\bibnamefont
  {Pasquier}}, \bibinfo {author} {\bibfnamefont {O.~V.}\ \bibnamefont
  {Yazyev}}, \ and\ \bibinfo {author} {\bibfnamefont {A.}~\bibnamefont {Kis}},\
  }\href {\doibase 10.1038/natrevmats.2017.33} {\bibfield  {journal} {\bibinfo
  {journal} {Nature Reviews Materials}\ }\textbf {\bibinfo {volume} {2}}
  (\bibinfo {year} {2017}),\ 10.1038/natrevmats.2017.33}\BibitemShut {NoStop}%
\bibitem [{\citenamefont {{Yue}}\ \emph {et~al.}(2012)\citenamefont {{Yue}},
  \citenamefont {{Kang}}, \citenamefont {{Shao}}, \citenamefont {{Zhang}},
  \citenamefont {{Chang}}, \citenamefont {{Wang}}, \citenamefont {{Qin}},\ and\
  \citenamefont {{Li}}}]{2012PhLA..376.1166Y}%
  \BibitemOpen
  \bibfield  {author} {\bibinfo {author} {\bibfnamefont {Q.}~\bibnamefont
  {{Yue}}}, \bibinfo {author} {\bibfnamefont {J.}~\bibnamefont {{Kang}}},
  \bibinfo {author} {\bibfnamefont {Z.}~\bibnamefont {{Shao}}}, \bibinfo
  {author} {\bibfnamefont {X.}~\bibnamefont {{Zhang}}}, \bibinfo {author}
  {\bibfnamefont {S.}~\bibnamefont {{Chang}}}, \bibinfo {author} {\bibfnamefont
  {G.}~\bibnamefont {{Wang}}}, \bibinfo {author} {\bibfnamefont
  {S.}~\bibnamefont {{Qin}}}, \ and\ \bibinfo {author} {\bibfnamefont
  {J.}~\bibnamefont {{Li}}},\ }\href {\doibase 10.1016/j.physleta.2012.02.029}
  {\bibfield  {journal} {\bibinfo  {journal} {Physics Letters A}\ }\textbf
  {\bibinfo {volume} {376}},\ \bibinfo {pages} {1166} (\bibinfo {year}
  {2012})}\BibitemShut {NoStop}%
\bibitem [{\citenamefont {Shi}\ \emph {et~al.}(2016)\citenamefont {Shi},
  \citenamefont {Tong}, \citenamefont {Zhou}, \citenamefont {Gong},
  \citenamefont {Zhang}, \citenamefont {Ji}, \citenamefont {Zhang},
  \citenamefont {Fang}, \citenamefont {Gu}, \citenamefont {Wang}, \citenamefont
  {Liu},\ and\ \citenamefont {Zhang}}]{photocatalytic_tmdc}%
  \BibitemOpen
  \bibfield  {author} {\bibinfo {author} {\bibfnamefont {J.}~\bibnamefont
  {Shi}}, \bibinfo {author} {\bibfnamefont {R.}~\bibnamefont {Tong}}, \bibinfo
  {author} {\bibfnamefont {X.}~\bibnamefont {Zhou}}, \bibinfo {author}
  {\bibfnamefont {Y.}~\bibnamefont {Gong}}, \bibinfo {author} {\bibfnamefont
  {Z.}~\bibnamefont {Zhang}}, \bibinfo {author} {\bibfnamefont
  {Q.}~\bibnamefont {Ji}}, \bibinfo {author} {\bibfnamefont {Y.}~\bibnamefont
  {Zhang}}, \bibinfo {author} {\bibfnamefont {Q.}~\bibnamefont {Fang}},
  \bibinfo {author} {\bibfnamefont {L.}~\bibnamefont {Gu}}, \bibinfo {author}
  {\bibfnamefont {X.}~\bibnamefont {Wang}}, \bibinfo {author} {\bibfnamefont
  {Z.}~\bibnamefont {Liu}}, \ and\ \bibinfo {author} {\bibfnamefont
  {Y.}~\bibnamefont {Zhang}},\ }\href {\doibase 10.1002/adma.201603174}
  {\bibfield  {journal} {\bibinfo  {journal} {Advanced Materials}\ }\textbf
  {\bibinfo {volume} {28}},\ \bibinfo {pages} {10664} (\bibinfo {year}
  {2016})}\BibitemShut {NoStop}%
\bibitem [{\citenamefont {Chhowalla}\ \emph {et~al.}(2013)\citenamefont
  {Chhowalla}, \citenamefont {Shin}, \citenamefont {Eda}, \citenamefont {Li},
  \citenamefont {Loh},\ and\ \citenamefont {Zhang}}]{chhowalla_tmdc_2013}%
  \BibitemOpen
  \bibfield  {author} {\bibinfo {author} {\bibfnamefont {M.}~\bibnamefont
  {Chhowalla}}, \bibinfo {author} {\bibfnamefont {H.~S.}\ \bibnamefont {Shin}},
  \bibinfo {author} {\bibfnamefont {G.}~\bibnamefont {Eda}}, \bibinfo {author}
  {\bibfnamefont {L.-J.}\ \bibnamefont {Li}}, \bibinfo {author} {\bibfnamefont
  {K.~P.}\ \bibnamefont {Loh}}, \ and\ \bibinfo {author} {\bibfnamefont
  {H.}~\bibnamefont {Zhang}},\ }\href {\doibase 10.1038/nchem.1589} {\bibfield
  {journal} {\bibinfo  {journal} {Nature Chemistry}\ }\textbf {\bibinfo
  {volume} {5}},\ \bibinfo {pages} {263} (\bibinfo {year} {2013})}\BibitemShut
  {NoStop}%
\bibitem [{\citenamefont {Klemm}(2012)}]{klemm_2012}%
  \BibitemOpen
  \bibfield  {author} {\bibinfo {author} {\bibfnamefont {R.~A.}\ \bibnamefont
  {Klemm}},\ }\enquote {\bibinfo {title} {Layered superconducting materials},}\
  in\ \href {\doibase 10.1093/acprof:oso/9780199593316.001.0001} {\emph
  {\bibinfo {booktitle} {Layered superconductors:Volume 1}}}\ (\bibinfo
  {publisher} {Oxford University Press},\ \bibinfo {year} {2012})\ pp.\
  \bibinfo {pages} {20--75}\BibitemShut {NoStop}%
\bibitem [{\citenamefont {Revolinsky}\ \emph {et~al.}(1965)\citenamefont
  {Revolinsky}, \citenamefont {Spiering},\ and\ \citenamefont
  {Beerntsen}}]{REVOLINSKY19651029}%
  \BibitemOpen
  \bibfield  {author} {\bibinfo {author} {\bibfnamefont {E.}~\bibnamefont
  {Revolinsky}}, \bibinfo {author} {\bibfnamefont {G.}~\bibnamefont
  {Spiering}}, \ and\ \bibinfo {author} {\bibfnamefont {D.}~\bibnamefont
  {Beerntsen}},\ }\href {\doibase https://doi.org/10.1016/0022-3697(65)90190-3}
  {\bibfield  {journal} {\bibinfo  {journal} {Journal of Physics and Chemistry
  of Solids}\ }\textbf {\bibinfo {volume} {26}},\ \bibinfo {pages} {1029 }
  (\bibinfo {year} {1965})}\BibitemShut {NoStop}%
\bibitem [{\citenamefont {Ugeda}\ \emph {et~al.}(2016)\citenamefont {Ugeda},
  \citenamefont {Bradley}, \citenamefont {Zhang}, \citenamefont {Onishi},
  \citenamefont {Chen}, \citenamefont {Ruan}, \citenamefont
  {Ojeda-Aristizabal}, \citenamefont {Ryu}, \citenamefont {Edmonds},
  \citenamefont {Tsai} \emph {et~al.}}]{ugeda2016characterization}%
  \BibitemOpen
  \bibfield  {author} {\bibinfo {author} {\bibfnamefont {M.~M.}\ \bibnamefont
  {Ugeda}}, \bibinfo {author} {\bibfnamefont {A.~J.}\ \bibnamefont {Bradley}},
  \bibinfo {author} {\bibfnamefont {Y.}~\bibnamefont {Zhang}}, \bibinfo
  {author} {\bibfnamefont {S.}~\bibnamefont {Onishi}}, \bibinfo {author}
  {\bibfnamefont {Y.}~\bibnamefont {Chen}}, \bibinfo {author} {\bibfnamefont
  {W.}~\bibnamefont {Ruan}}, \bibinfo {author} {\bibfnamefont {C.}~\bibnamefont
  {Ojeda-Aristizabal}}, \bibinfo {author} {\bibfnamefont {H.}~\bibnamefont
  {Ryu}}, \bibinfo {author} {\bibfnamefont {M.~T.}\ \bibnamefont {Edmonds}},
  \bibinfo {author} {\bibfnamefont {H.-Z.}\ \bibnamefont {Tsai}},  \emph
  {et~al.},\ }\href@noop {} {\bibfield  {journal} {\bibinfo  {journal} {Nature
  Physics}\ }\textbf {\bibinfo {volume} {12}},\ \bibinfo {pages} {92} (\bibinfo
  {year} {2016})}\BibitemShut {NoStop}%
\bibitem [{\citenamefont {Lian}\ \emph {et~al.}(2018)\citenamefont {Lian},
  \citenamefont {Si},\ and\ \citenamefont {Duan}}]{Lian_nbse2}%
  \BibitemOpen
  \bibfield  {author} {\bibinfo {author} {\bibfnamefont {C.-S.}\ \bibnamefont
  {Lian}}, \bibinfo {author} {\bibfnamefont {C.}~\bibnamefont {Si}}, \ and\
  \bibinfo {author} {\bibfnamefont {W.}~\bibnamefont {Duan}},\ }\href {\doibase
  10.1021/acs.nanolett.8b00237} {\bibfield  {journal} {\bibinfo  {journal}
  {Nano Letters}\ }\textbf {\bibinfo {volume} {18}},\ \bibinfo {pages} {2924}
  (\bibinfo {year} {2018})},\ \bibinfo {note} {pMID: 29652158}\BibitemShut
  {NoStop}%
\bibitem [{\citenamefont {Zheng}\ and\ \citenamefont
  {Feng}(2019)}]{PhysRevB.99.161119}%
  \BibitemOpen
  \bibfield  {author} {\bibinfo {author} {\bibfnamefont {F.}~\bibnamefont
  {Zheng}}\ and\ \bibinfo {author} {\bibfnamefont {J.}~\bibnamefont {Feng}},\
  }\href@noop {} {\bibfield  {journal} {\bibinfo  {journal} {Phys. Rev. B}\
  }\textbf {\bibinfo {volume} {99}},\ \bibinfo {pages} {161119} (\bibinfo
  {year} {2019})}\BibitemShut {NoStop}%
\bibitem [{\citenamefont {{Somoano}}\ \emph {et~al.}(1973)\citenamefont
  {{Somoano}}, \citenamefont {{Hadek}},\ and\ \citenamefont
  {{Rembaum}}}]{1973JChPh..58..697S}%
  \BibitemOpen
  \bibfield  {author} {\bibinfo {author} {\bibfnamefont {R.~B.}\ \bibnamefont
  {{Somoano}}}, \bibinfo {author} {\bibfnamefont {V.}~\bibnamefont {{Hadek}}},
  \ and\ \bibinfo {author} {\bibfnamefont {A.}~\bibnamefont {{Rembaum}}},\
  }\href {\doibase 10.1063/1.1679256} {\bibfield  {journal} {\bibinfo
  {journal} {\jcp}\ }\textbf {\bibinfo {volume} {58}},\ \bibinfo {pages} {697}
  (\bibinfo {year} {1973})}\BibitemShut {NoStop}%
\bibitem [{\citenamefont {Woollam}\ and\ \citenamefont
  {Somoano}(1977)}]{WOOLLAM1977289}%
  \BibitemOpen
  \bibfield  {author} {\bibinfo {author} {\bibfnamefont {J.~A.}\ \bibnamefont
  {Woollam}}\ and\ \bibinfo {author} {\bibfnamefont {R.~B.}\ \bibnamefont
  {Somoano}},\ }\href {\doibase https://doi.org/10.1016/0025-5416(77)90048-9}
  {\bibfield  {journal} {\bibinfo  {journal} {Materials Science and
  Engineering}\ }\textbf {\bibinfo {volume} {31}},\ \bibinfo {pages} {289 }
  (\bibinfo {year} {1977})},\ \bibinfo {note} {proceedings of the Franco
  American Conference on Intercalation Compounds of Graphite}\BibitemShut
  {NoStop}%
\bibitem [{\citenamefont {Sipos}\ \emph {et~al.}(2008)\citenamefont {Sipos},
  \citenamefont {Kusmartseva}, \citenamefont {Akrap}, \citenamefont {Berger},
  \citenamefont {Forró},\ and\ \citenamefont {Tutiš}}]{sipos_Tas2_2008}%
  \BibitemOpen
  \bibfield  {author} {\bibinfo {author} {\bibfnamefont {B.}~\bibnamefont
  {Sipos}}, \bibinfo {author} {\bibfnamefont {A.~F.}\ \bibnamefont
  {Kusmartseva}}, \bibinfo {author} {\bibfnamefont {A.}~\bibnamefont {Akrap}},
  \bibinfo {author} {\bibfnamefont {H.}~\bibnamefont {Berger}}, \bibinfo
  {author} {\bibfnamefont {L.}~\bibnamefont {Forró}}, \ and\ \bibinfo {author}
  {\bibfnamefont {E.}~\bibnamefont {Tutiš}},\ }\href {\doibase
  10.1038/nmat2318} {\bibfield  {journal} {\bibinfo  {journal} {Nature
  Materials}\ }\textbf {\bibinfo {volume} {7}},\ \bibinfo {pages} {960}
  (\bibinfo {year} {2008})}\BibitemShut {NoStop}%
\bibitem [{\citenamefont {Ye}\ \emph {et~al.}(2012)\citenamefont {Ye},
  \citenamefont {Zhang}, \citenamefont {Akashi}, \citenamefont {Bahramy},
  \citenamefont {Arita},\ and\ \citenamefont {Iwasa}}]{Ye1193}%
  \BibitemOpen
  \bibfield  {author} {\bibinfo {author} {\bibfnamefont {J.~T.}\ \bibnamefont
  {Ye}}, \bibinfo {author} {\bibfnamefont {Y.~J.}\ \bibnamefont {Zhang}},
  \bibinfo {author} {\bibfnamefont {R.}~\bibnamefont {Akashi}}, \bibinfo
  {author} {\bibfnamefont {M.~S.}\ \bibnamefont {Bahramy}}, \bibinfo {author}
  {\bibfnamefont {R.}~\bibnamefont {Arita}}, \ and\ \bibinfo {author}
  {\bibfnamefont {Y.}~\bibnamefont {Iwasa}},\ }\href {\doibase
  10.1126/science.1228006} {\bibfield  {journal} {\bibinfo  {journal}
  {Science}\ }\textbf {\bibinfo {volume} {338}},\ \bibinfo {pages} {1193}
  (\bibinfo {year} {2012})}\BibitemShut {NoStop}%
\bibitem [{\citenamefont {Soluyanov}\ \emph {et~al.}(2015)\citenamefont
  {Soluyanov}, \citenamefont {Gresch}, \citenamefont {Wang}, \citenamefont
  {Wu}, \citenamefont {Troyer}, \citenamefont {Dai},\ and\ \citenamefont
  {Bernevig}}]{soluyanov_2015}%
  \BibitemOpen
  \bibfield  {author} {\bibinfo {author} {\bibfnamefont {A.~A.}\ \bibnamefont
  {Soluyanov}}, \bibinfo {author} {\bibfnamefont {D.}~\bibnamefont {Gresch}},
  \bibinfo {author} {\bibfnamefont {Z.}~\bibnamefont {Wang}}, \bibinfo {author}
  {\bibfnamefont {Q.}~\bibnamefont {Wu}}, \bibinfo {author} {\bibfnamefont
  {M.}~\bibnamefont {Troyer}}, \bibinfo {author} {\bibfnamefont
  {X.}~\bibnamefont {Dai}}, \ and\ \bibinfo {author} {\bibfnamefont {B.~A.}\
  \bibnamefont {Bernevig}},\ }\href {\doibase 10.1038/nature15768} {\bibfield
  {journal} {\bibinfo  {journal} {Nature}\ }\textbf {\bibinfo {volume} {527}},\
  \bibinfo {pages} {495} (\bibinfo {year} {2015})}\BibitemShut {NoStop}%
\bibitem [{\citenamefont {Sun}\ \emph {et~al.}(2015)\citenamefont {Sun},
  \citenamefont {Wu}, \citenamefont {Ali}, \citenamefont {Felser},\ and\
  \citenamefont {Yan}}]{PhysRevB.92.161107}%
  \BibitemOpen
  \bibfield  {author} {\bibinfo {author} {\bibfnamefont {Y.}~\bibnamefont
  {Sun}}, \bibinfo {author} {\bibfnamefont {S.-C.}\ \bibnamefont {Wu}},
  \bibinfo {author} {\bibfnamefont {M.~N.}\ \bibnamefont {Ali}}, \bibinfo
  {author} {\bibfnamefont {C.}~\bibnamefont {Felser}}, \ and\ \bibinfo {author}
  {\bibfnamefont {B.}~\bibnamefont {Yan}},\ }\href {\doibase
  10.1103/PhysRevB.92.161107} {\bibfield  {journal} {\bibinfo  {journal} {Phys.
  Rev. B}\ }\textbf {\bibinfo {volume} {92}},\ \bibinfo {pages} {161107}
  (\bibinfo {year} {2015})}\BibitemShut {NoStop}%
\bibitem [{\citenamefont {Pletikosi\ifmmode~\acute{c}\else \'{c}\fi{}}\ \emph
  {et~al.}(2014)\citenamefont {Pletikosi\ifmmode~\acute{c}\else \'{c}\fi{}},
  \citenamefont {Ali}, \citenamefont {Fedorov}, \citenamefont {Cava},\ and\
  \citenamefont {Valla}}]{PhysRevLett.113.216601}%
  \BibitemOpen
  \bibfield  {author} {\bibinfo {author} {\bibfnamefont {I.}~\bibnamefont
  {Pletikosi\ifmmode~\acute{c}\else \'{c}\fi{}}}, \bibinfo {author}
  {\bibfnamefont {M.~N.}\ \bibnamefont {Ali}}, \bibinfo {author} {\bibfnamefont
  {A.~V.}\ \bibnamefont {Fedorov}}, \bibinfo {author} {\bibfnamefont {R.~J.}\
  \bibnamefont {Cava}}, \ and\ \bibinfo {author} {\bibfnamefont
  {T.}~\bibnamefont {Valla}},\ }\href {\doibase 10.1103/PhysRevLett.113.216601}
  {\bibfield  {journal} {\bibinfo  {journal} {Phys. Rev. Lett.}\ }\textbf
  {\bibinfo {volume} {113}},\ \bibinfo {pages} {216601} (\bibinfo {year}
  {2014})}\BibitemShut {NoStop}%
\bibitem [{\citenamefont {Ali}\ \emph {et~al.}(2014)\citenamefont {Ali},
  \citenamefont {Xiong}, \citenamefont {Flynn}, \citenamefont {Tao},
  \citenamefont {Gibson}, \citenamefont {Schoop}, \citenamefont {Liang},
  \citenamefont {Haldolaarachchige}, \citenamefont {Hirschberger},\ and\
  \citenamefont {Ong}}]{wte2_mag_resistance}%
  \BibitemOpen
  \bibfield  {author} {\bibinfo {author} {\bibfnamefont {M.~N.}\ \bibnamefont
  {Ali}}, \bibinfo {author} {\bibfnamefont {J.}~\bibnamefont {Xiong}}, \bibinfo
  {author} {\bibfnamefont {S.}~\bibnamefont {Flynn}}, \bibinfo {author}
  {\bibfnamefont {J.}~\bibnamefont {Tao}}, \bibinfo {author} {\bibfnamefont
  {Q.~D.}\ \bibnamefont {Gibson}}, \bibinfo {author} {\bibfnamefont {L.~M.}\
  \bibnamefont {Schoop}}, \bibinfo {author} {\bibfnamefont {T.}~\bibnamefont
  {Liang}}, \bibinfo {author} {\bibfnamefont {N.}~\bibnamefont
  {Haldolaarachchige}}, \bibinfo {author} {\bibfnamefont {M.}~\bibnamefont
  {Hirschberger}}, \ and\ \bibinfo {author} {\bibfnamefont {N.~P. e.~a.}\
  \bibnamefont {Ong}},\ }\href {\doibase 10.1038/nature13763} {\bibfield
  {journal} {\bibinfo  {journal} {Nature}\ }\textbf {\bibinfo {volume} {514}},\
  \bibinfo {pages} {205} (\bibinfo {year} {2014})}\BibitemShut {NoStop}%
\bibitem [{\citenamefont {Keum}\ \emph {et~al.}(2015)\citenamefont {Keum},
  \citenamefont {Cho}, \citenamefont {Kim}, \citenamefont {Choe}, \citenamefont
  {Sung}, \citenamefont {Kan}, \citenamefont {Kang}, \citenamefont {Hwang},
  \citenamefont {Kim},\ and\ \citenamefont {Yang}}]{keum_mote2_2015}%
  \BibitemOpen
  \bibfield  {author} {\bibinfo {author} {\bibfnamefont {D.~H.}\ \bibnamefont
  {Keum}}, \bibinfo {author} {\bibfnamefont {S.}~\bibnamefont {Cho}}, \bibinfo
  {author} {\bibfnamefont {J.~H.}\ \bibnamefont {Kim}}, \bibinfo {author}
  {\bibfnamefont {D.-H.}\ \bibnamefont {Choe}}, \bibinfo {author}
  {\bibfnamefont {H.-J.}\ \bibnamefont {Sung}}, \bibinfo {author}
  {\bibfnamefont {M.}~\bibnamefont {Kan}}, \bibinfo {author} {\bibfnamefont
  {H.}~\bibnamefont {Kang}}, \bibinfo {author} {\bibfnamefont {J.-Y.}\
  \bibnamefont {Hwang}}, \bibinfo {author} {\bibfnamefont {S.~W.}\ \bibnamefont
  {Kim}}, \ and\ \bibinfo {author} {\bibfnamefont {H.~e.~a.}\ \bibnamefont
  {Yang}},\ }\href {\doibase 10.1038/nphys3314} {\bibfield  {journal} {\bibinfo
   {journal} {Nature Physics}\ }\textbf {\bibinfo {volume} {11}},\ \bibinfo
  {pages} {482} (\bibinfo {year} {2015})}\BibitemShut {NoStop}%
\bibitem [{\citenamefont {Qian}\ \emph {et~al.}(2014)\citenamefont {Qian},
  \citenamefont {Liu}, \citenamefont {Fu},\ and\ \citenamefont
  {Li}}]{Qian1344}%
  \BibitemOpen
  \bibfield  {author} {\bibinfo {author} {\bibfnamefont {X.}~\bibnamefont
  {Qian}}, \bibinfo {author} {\bibfnamefont {J.}~\bibnamefont {Liu}}, \bibinfo
  {author} {\bibfnamefont {L.}~\bibnamefont {Fu}}, \ and\ \bibinfo {author}
  {\bibfnamefont {J.}~\bibnamefont {Li}},\ }\href {\doibase
  10.1126/science.1256815} {\bibfield  {journal} {\bibinfo  {journal}
  {Science}\ }\textbf {\bibinfo {volume} {346}},\ \bibinfo {pages} {1344}
  (\bibinfo {year} {2014})}\BibitemShut {NoStop}%
\bibitem [{\citenamefont {Tang}\ \emph {et~al.}(2017)\citenamefont {Tang},
  \citenamefont {Zhang}, \citenamefont {Wong}, \citenamefont {Pedramrazi},
  \citenamefont {Tsai}, \citenamefont {Jia}, \citenamefont {Moritz},
  \citenamefont {Claassen}, \citenamefont {Ryu}, \citenamefont {Kahn} \emph
  {et~al.}}]{tang2017quantum}%
  \BibitemOpen
  \bibfield  {author} {\bibinfo {author} {\bibfnamefont {S.}~\bibnamefont
  {Tang}}, \bibinfo {author} {\bibfnamefont {C.}~\bibnamefont {Zhang}},
  \bibinfo {author} {\bibfnamefont {D.}~\bibnamefont {Wong}}, \bibinfo {author}
  {\bibfnamefont {Z.}~\bibnamefont {Pedramrazi}}, \bibinfo {author}
  {\bibfnamefont {H.-Z.}\ \bibnamefont {Tsai}}, \bibinfo {author}
  {\bibfnamefont {C.}~\bibnamefont {Jia}}, \bibinfo {author} {\bibfnamefont
  {B.}~\bibnamefont {Moritz}}, \bibinfo {author} {\bibfnamefont
  {M.}~\bibnamefont {Claassen}}, \bibinfo {author} {\bibfnamefont
  {H.}~\bibnamefont {Ryu}}, \bibinfo {author} {\bibfnamefont {S.}~\bibnamefont
  {Kahn}},  \emph {et~al.},\ }\href@noop {} {\bibfield  {journal} {\bibinfo
  {journal} {Nature Physics}\ }\textbf {\bibinfo {volume} {13}},\ \bibinfo
  {pages} {683} (\bibinfo {year} {2017})}\BibitemShut {NoStop}%
\bibitem [{\citenamefont {Fei}\ \emph {et~al.}(2017)\citenamefont {Fei},
  \citenamefont {Palomaki}, \citenamefont {Wu}, \citenamefont {Zhao},
  \citenamefont {Cai}, \citenamefont {Sun}, \citenamefont {Nguyen},
  \citenamefont {Finney}, \citenamefont {Xu},\ and\ \citenamefont
  {Cobden}}]{fei2017edge}%
  \BibitemOpen
  \bibfield  {author} {\bibinfo {author} {\bibfnamefont {Z.}~\bibnamefont
  {Fei}}, \bibinfo {author} {\bibfnamefont {T.}~\bibnamefont {Palomaki}},
  \bibinfo {author} {\bibfnamefont {S.}~\bibnamefont {Wu}}, \bibinfo {author}
  {\bibfnamefont {W.}~\bibnamefont {Zhao}}, \bibinfo {author} {\bibfnamefont
  {X.}~\bibnamefont {Cai}}, \bibinfo {author} {\bibfnamefont {B.}~\bibnamefont
  {Sun}}, \bibinfo {author} {\bibfnamefont {P.}~\bibnamefont {Nguyen}},
  \bibinfo {author} {\bibfnamefont {J.}~\bibnamefont {Finney}}, \bibinfo
  {author} {\bibfnamefont {X.}~\bibnamefont {Xu}}, \ and\ \bibinfo {author}
  {\bibfnamefont {D.~H.}\ \bibnamefont {Cobden}},\ }\href@noop {} {\bibfield
  {journal} {\bibinfo  {journal} {Nature Physics}\ }\textbf {\bibinfo {volume}
  {13}},\ \bibinfo {pages} {677} (\bibinfo {year} {2017})}\BibitemShut
  {NoStop}%
\bibitem [{\citenamefont {Wu}\ \emph {et~al.}(2018{\natexlab{a}})\citenamefont
  {Wu}, \citenamefont {Fatemi}, \citenamefont {Gibson}, \citenamefont
  {Watanabe}, \citenamefont {Taniguchi}, \citenamefont {Cava},\ and\
  \citenamefont {Jarillo-Herrero}}]{wu2018observation}%
  \BibitemOpen
  \bibfield  {author} {\bibinfo {author} {\bibfnamefont {S.}~\bibnamefont
  {Wu}}, \bibinfo {author} {\bibfnamefont {V.}~\bibnamefont {Fatemi}}, \bibinfo
  {author} {\bibfnamefont {Q.~D.}\ \bibnamefont {Gibson}}, \bibinfo {author}
  {\bibfnamefont {K.}~\bibnamefont {Watanabe}}, \bibinfo {author}
  {\bibfnamefont {T.}~\bibnamefont {Taniguchi}}, \bibinfo {author}
  {\bibfnamefont {R.~J.}\ \bibnamefont {Cava}}, \ and\ \bibinfo {author}
  {\bibfnamefont {P.}~\bibnamefont {Jarillo-Herrero}},\ }\href@noop {}
  {\bibfield  {journal} {\bibinfo  {journal} {Science}\ }\textbf {\bibinfo
  {volume} {359}},\ \bibinfo {pages} {76} (\bibinfo {year}
  {2018}{\natexlab{a}})}\BibitemShut {NoStop}%
\bibitem [{\citenamefont {Zhou}\ \emph {et~al.}(2016)\citenamefont {Zhou},
  \citenamefont {Yuan}, \citenamefont {Jiang},\ and\ \citenamefont
  {Law}}]{PhysRevB.93.180501}%
  \BibitemOpen
  \bibfield  {author} {\bibinfo {author} {\bibfnamefont {B.~T.}\ \bibnamefont
  {Zhou}}, \bibinfo {author} {\bibfnamefont {N.~F.~Q.}\ \bibnamefont {Yuan}},
  \bibinfo {author} {\bibfnamefont {H.-L.}\ \bibnamefont {Jiang}}, \ and\
  \bibinfo {author} {\bibfnamefont {K.~T.}\ \bibnamefont {Law}},\ }\href
  {\doibase 10.1103/PhysRevB.93.180501} {\bibfield  {journal} {\bibinfo
  {journal} {Phys. Rev. B}\ }\textbf {\bibinfo {volume} {93}},\ \bibinfo
  {pages} {180501} (\bibinfo {year} {2016})}\BibitemShut {NoStop}%
\bibitem [{\citenamefont {Fang}\ \emph {et~al.}(2019)\citenamefont {Fang},
  \citenamefont {Pan}, \citenamefont {Zhang}, \citenamefont {Wang},
  \citenamefont {Hirose}, \citenamefont {Terashima}, \citenamefont {Uji},
  \citenamefont {Yuan}, \citenamefont {Li}, \citenamefont {Tian}, \citenamefont
  {Xue}, \citenamefont {Ma}, \citenamefont {Zhao}, \citenamefont {Xue},
  \citenamefont {Mu}, \citenamefont {Zhang},\ and\ \citenamefont
  {Huang}}]{fang2019discovery}%
  \BibitemOpen
  \bibfield  {author} {\bibinfo {author} {\bibfnamefont {Y.}~\bibnamefont
  {Fang}}, \bibinfo {author} {\bibfnamefont {J.}~\bibnamefont {Pan}}, \bibinfo
  {author} {\bibfnamefont {D.}~\bibnamefont {Zhang}}, \bibinfo {author}
  {\bibfnamefont {D.}~\bibnamefont {Wang}}, \bibinfo {author} {\bibfnamefont
  {H.~T.}\ \bibnamefont {Hirose}}, \bibinfo {author} {\bibfnamefont
  {T.}~\bibnamefont {Terashima}}, \bibinfo {author} {\bibfnamefont
  {S.}~\bibnamefont {Uji}}, \bibinfo {author} {\bibfnamefont {Y.}~\bibnamefont
  {Yuan}}, \bibinfo {author} {\bibfnamefont {W.}~\bibnamefont {Li}}, \bibinfo
  {author} {\bibfnamefont {Z.}~\bibnamefont {Tian}}, \bibinfo {author}
  {\bibfnamefont {J.}~\bibnamefont {Xue}}, \bibinfo {author} {\bibfnamefont
  {Y.}~\bibnamefont {Ma}}, \bibinfo {author} {\bibfnamefont {W.}~\bibnamefont
  {Zhao}}, \bibinfo {author} {\bibfnamefont {Q.}~\bibnamefont {Xue}}, \bibinfo
  {author} {\bibfnamefont {G.}~\bibnamefont {Mu}}, \bibinfo {author}
  {\bibfnamefont {H.}~\bibnamefont {Zhang}}, \ and\ \bibinfo {author}
  {\bibfnamefont {F.}~\bibnamefont {Huang}},\ }\href {\doibase
  10.1002/adma.201901942} {\bibfield  {journal} {\bibinfo  {journal} {Advanced
  Materials}\ }\textbf {\bibinfo {volume} {31}} (\bibinfo {year} {2019}),\
  10.1002/adma.201901942}\BibitemShut {NoStop}%
\bibitem [{\citenamefont {Fang}\ \emph {et~al.}(2020)\citenamefont {Fang},
  \citenamefont {Wang}, \citenamefont {Zhao},\ and\ \citenamefont
  {Huang}}]{Fang_2020}%
  \BibitemOpen
  \bibfield  {author} {\bibinfo {author} {\bibfnamefont {Y.~Q.}\ \bibnamefont
  {Fang}}, \bibinfo {author} {\bibfnamefont {D.}~\bibnamefont {Wang}}, \bibinfo
  {author} {\bibfnamefont {W.}~\bibnamefont {Zhao}}, \ and\ \bibinfo {author}
  {\bibfnamefont {F.~Q.}\ \bibnamefont {Huang}},\ }\href {\doibase
  10.1209/0295-5075/131/10005} {\bibfield  {journal} {\bibinfo  {journal}
  {{EPL} (Europhysics Letters)}\ }\textbf {\bibinfo {volume} {131}},\ \bibinfo
  {pages} {10005} (\bibinfo {year} {2020})}\BibitemShut {NoStop}%
\bibitem [{\citenamefont {Guguchia}\ \emph {et~al.}(2019)\citenamefont
  {Guguchia}, \citenamefont {Gawryluk}, \citenamefont {Brzezinska},
  \citenamefont {Tsirkin}, \citenamefont {Khasanov}, \citenamefont
  {Pomjakushina}, \citenamefont {von Rohr}, \citenamefont {Verezhak},
  \citenamefont {Hasan}, \citenamefont {Neupert} \emph
  {et~al.}}]{guguchia2019nodeless}%
  \BibitemOpen
  \bibfield  {author} {\bibinfo {author} {\bibfnamefont {Z.}~\bibnamefont
  {Guguchia}}, \bibinfo {author} {\bibfnamefont {D.~J.}\ \bibnamefont
  {Gawryluk}}, \bibinfo {author} {\bibfnamefont {M.}~\bibnamefont
  {Brzezinska}}, \bibinfo {author} {\bibfnamefont {S.~S.}\ \bibnamefont
  {Tsirkin}}, \bibinfo {author} {\bibfnamefont {R.}~\bibnamefont {Khasanov}},
  \bibinfo {author} {\bibfnamefont {E.}~\bibnamefont {Pomjakushina}}, \bibinfo
  {author} {\bibfnamefont {F.~O.}\ \bibnamefont {von Rohr}}, \bibinfo {author}
  {\bibfnamefont {J.~A.}\ \bibnamefont {Verezhak}}, \bibinfo {author}
  {\bibfnamefont {M.~Z.}\ \bibnamefont {Hasan}}, \bibinfo {author}
  {\bibfnamefont {T.}~\bibnamefont {Neupert}},  \emph {et~al.},\ }\href@noop {}
  {\bibfield  {journal} {\bibinfo  {journal} {npj Quantum Materials}\ }\textbf
  {\bibinfo {volume} {4}},\ \bibinfo {pages} {1} (\bibinfo {year}
  {2019})}\BibitemShut {NoStop}%
\bibitem [{\citenamefont {Yuan}\ \emph {et~al.}(2019)\citenamefont {Yuan},
  \citenamefont {Pan}, \citenamefont {Wang}, \citenamefont {Fang},
  \citenamefont {Song}, \citenamefont {Wang}, \citenamefont {He}, \citenamefont
  {Ma}, \citenamefont {Zhang}, \citenamefont {Huang} \emph
  {et~al.}}]{yuan2019evidence}%
  \BibitemOpen
  \bibfield  {author} {\bibinfo {author} {\bibfnamefont {Y.}~\bibnamefont
  {Yuan}}, \bibinfo {author} {\bibfnamefont {J.}~\bibnamefont {Pan}}, \bibinfo
  {author} {\bibfnamefont {X.}~\bibnamefont {Wang}}, \bibinfo {author}
  {\bibfnamefont {Y.}~\bibnamefont {Fang}}, \bibinfo {author} {\bibfnamefont
  {C.}~\bibnamefont {Song}}, \bibinfo {author} {\bibfnamefont {L.}~\bibnamefont
  {Wang}}, \bibinfo {author} {\bibfnamefont {K.}~\bibnamefont {He}}, \bibinfo
  {author} {\bibfnamefont {X.}~\bibnamefont {Ma}}, \bibinfo {author}
  {\bibfnamefont {H.}~\bibnamefont {Zhang}}, \bibinfo {author} {\bibfnamefont
  {F.}~\bibnamefont {Huang}},  \emph {et~al.},\ }\href@noop {} {\bibfield
  {journal} {\bibinfo  {journal} {Nature Physics}\ }\textbf {\bibinfo {volume}
  {15}},\ \bibinfo {pages} {1046} (\bibinfo {year} {2019})}\BibitemShut
  {NoStop}%
\bibitem [{\citenamefont {Wang}\ \emph {et~al.}(2020)\citenamefont {Wang},
  \citenamefont {Fang}, \citenamefont {Huang}, \citenamefont {Cheng},
  \citenamefont {Ni}, \citenamefont {Pan}, \citenamefont {Xu}, \citenamefont
  {Huang},\ and\ \citenamefont {Li}}]{PhysRevB.102.024523}%
  \BibitemOpen
  \bibfield  {author} {\bibinfo {author} {\bibfnamefont {L.~S.}\ \bibnamefont
  {Wang}}, \bibinfo {author} {\bibfnamefont {Y.~Q.}\ \bibnamefont {Fang}},
  \bibinfo {author} {\bibfnamefont {Y.~Y.}\ \bibnamefont {Huang}}, \bibinfo
  {author} {\bibfnamefont {E.~J.}\ \bibnamefont {Cheng}}, \bibinfo {author}
  {\bibfnamefont {J.~M.}\ \bibnamefont {Ni}}, \bibinfo {author} {\bibfnamefont
  {B.~L.}\ \bibnamefont {Pan}}, \bibinfo {author} {\bibfnamefont
  {Y.}~\bibnamefont {Xu}}, \bibinfo {author} {\bibfnamefont {F.~Q.}\
  \bibnamefont {Huang}}, \ and\ \bibinfo {author} {\bibfnamefont {S.~Y.}\
  \bibnamefont {Li}},\ }\href {\doibase 10.1103/PhysRevB.102.024523} {\bibfield
   {journal} {\bibinfo  {journal} {Phys. Rev. B}\ }\textbf {\bibinfo {volume}
  {102}},\ \bibinfo {pages} {024523} (\bibinfo {year} {2020})}\BibitemShut
  {NoStop}%
\bibitem [{\citenamefont {Giannozzi}\ \emph {et~al.}(2017)\citenamefont
  {Giannozzi}, \citenamefont {Andreussi}, \citenamefont {Brumme}, \citenamefont
  {Bunau}, \citenamefont {Nardelli}, \citenamefont {Calandra}, \citenamefont
  {Car}, \citenamefont {Cavazzoni}, \citenamefont {Ceresoli}, \citenamefont
  {Cococcioni}, \citenamefont {Colonna}, \citenamefont {Carnimeo},
  \citenamefont {Corso}, \citenamefont {de~Gironcoli}, \citenamefont {Delugas},
  \citenamefont {Jr}, \citenamefont {Ferretti}, \citenamefont {Floris},
  \citenamefont {Fratesi}, \citenamefont {Fugallo}, \citenamefont {Gebauer},
  \citenamefont {Gerstmann}, \citenamefont {Giustino}, \citenamefont {Gorni},
  \citenamefont {Jia}, \citenamefont {Kawamura}, \citenamefont {Ko},
  \citenamefont {Kokalj}, \citenamefont {Küçükbenli}, \citenamefont
  {Lazzeri}, \citenamefont {Marsili}, \citenamefont {Marzari}, \citenamefont
  {Mauri}, \citenamefont {Nguyen}, \citenamefont {Nguyen}, \citenamefont {de-la
  Roza}, \citenamefont {Paulatto}, \citenamefont {Poncé}, \citenamefont
  {Rocca}, \citenamefont {Sabatini}, \citenamefont {Santra}, \citenamefont
  {Schlipf}, \citenamefont {Seitsonen}, \citenamefont {Smogunov}, \citenamefont
  {Timrov}, \citenamefont {Thonhauser}, \citenamefont {Umari}, \citenamefont
  {Vast}, \citenamefont {Wu},\ and\ \citenamefont {Baroni}}]{QE-2017}%
  \BibitemOpen
  \bibfield  {author} {\bibinfo {author} {\bibfnamefont {P.}~\bibnamefont
  {Giannozzi}}, \bibinfo {author} {\bibfnamefont {O.}~\bibnamefont
  {Andreussi}}, \bibinfo {author} {\bibfnamefont {T.}~\bibnamefont {Brumme}},
  \bibinfo {author} {\bibfnamefont {O.}~\bibnamefont {Bunau}}, \bibinfo
  {author} {\bibfnamefont {M.~B.}\ \bibnamefont {Nardelli}}, \bibinfo {author}
  {\bibfnamefont {M.}~\bibnamefont {Calandra}}, \bibinfo {author}
  {\bibfnamefont {R.}~\bibnamefont {Car}}, \bibinfo {author} {\bibfnamefont
  {C.}~\bibnamefont {Cavazzoni}}, \bibinfo {author} {\bibfnamefont
  {D.}~\bibnamefont {Ceresoli}}, \bibinfo {author} {\bibfnamefont
  {M.}~\bibnamefont {Cococcioni}}, \bibinfo {author} {\bibfnamefont
  {N.}~\bibnamefont {Colonna}}, \bibinfo {author} {\bibfnamefont
  {I.}~\bibnamefont {Carnimeo}}, \bibinfo {author} {\bibfnamefont {A.~D.}\
  \bibnamefont {Corso}}, \bibinfo {author} {\bibfnamefont {S.}~\bibnamefont
  {de~Gironcoli}}, \bibinfo {author} {\bibfnamefont {P.}~\bibnamefont
  {Delugas}}, \bibinfo {author} {\bibfnamefont {R.~A.~D.}\ \bibnamefont {Jr}},
  \bibinfo {author} {\bibfnamefont {A.}~\bibnamefont {Ferretti}}, \bibinfo
  {author} {\bibfnamefont {A.}~\bibnamefont {Floris}}, \bibinfo {author}
  {\bibfnamefont {G.}~\bibnamefont {Fratesi}}, \bibinfo {author} {\bibfnamefont
  {G.}~\bibnamefont {Fugallo}}, \bibinfo {author} {\bibfnamefont
  {R.}~\bibnamefont {Gebauer}}, \bibinfo {author} {\bibfnamefont
  {U.}~\bibnamefont {Gerstmann}}, \bibinfo {author} {\bibfnamefont
  {F.}~\bibnamefont {Giustino}}, \bibinfo {author} {\bibfnamefont
  {T.}~\bibnamefont {Gorni}}, \bibinfo {author} {\bibfnamefont
  {J.}~\bibnamefont {Jia}}, \bibinfo {author} {\bibfnamefont {M.}~\bibnamefont
  {Kawamura}}, \bibinfo {author} {\bibfnamefont {H.-Y.}\ \bibnamefont {Ko}},
  \bibinfo {author} {\bibfnamefont {A.}~\bibnamefont {Kokalj}}, \bibinfo
  {author} {\bibfnamefont {E.}~\bibnamefont {Küçükbenli}}, \bibinfo {author}
  {\bibfnamefont {M.}~\bibnamefont {Lazzeri}}, \bibinfo {author} {\bibfnamefont
  {M.}~\bibnamefont {Marsili}}, \bibinfo {author} {\bibfnamefont
  {N.}~\bibnamefont {Marzari}}, \bibinfo {author} {\bibfnamefont
  {F.}~\bibnamefont {Mauri}}, \bibinfo {author} {\bibfnamefont {N.~L.}\
  \bibnamefont {Nguyen}}, \bibinfo {author} {\bibfnamefont {H.-V.}\
  \bibnamefont {Nguyen}}, \bibinfo {author} {\bibfnamefont {A.~O.}\
  \bibnamefont {de-la Roza}}, \bibinfo {author} {\bibfnamefont
  {L.}~\bibnamefont {Paulatto}}, \bibinfo {author} {\bibfnamefont
  {S.}~\bibnamefont {Poncé}}, \bibinfo {author} {\bibfnamefont
  {D.}~\bibnamefont {Rocca}}, \bibinfo {author} {\bibfnamefont
  {R.}~\bibnamefont {Sabatini}}, \bibinfo {author} {\bibfnamefont
  {B.}~\bibnamefont {Santra}}, \bibinfo {author} {\bibfnamefont
  {M.}~\bibnamefont {Schlipf}}, \bibinfo {author} {\bibfnamefont {A.~P.}\
  \bibnamefont {Seitsonen}}, \bibinfo {author} {\bibfnamefont {A.}~\bibnamefont
  {Smogunov}}, \bibinfo {author} {\bibfnamefont {I.}~\bibnamefont {Timrov}},
  \bibinfo {author} {\bibfnamefont {T.}~\bibnamefont {Thonhauser}}, \bibinfo
  {author} {\bibfnamefont {P.}~\bibnamefont {Umari}}, \bibinfo {author}
  {\bibfnamefont {N.}~\bibnamefont {Vast}}, \bibinfo {author} {\bibfnamefont
  {X.}~\bibnamefont {Wu}}, \ and\ \bibinfo {author} {\bibfnamefont
  {S.}~\bibnamefont {Baroni}},\ }\href@noop {} {\bibfield  {journal} {\bibinfo
  {journal} {Journal of Physics: Condensed Matter}\ }\textbf {\bibinfo {volume}
  {29}},\ \bibinfo {pages} {465901} (\bibinfo {year} {2017})}\BibitemShut
  {NoStop}%
\bibitem [{\citenamefont {Giannozzi}\ \emph {et~al.}(2009)\citenamefont
  {Giannozzi}, \citenamefont {Baroni}, \citenamefont {Bonini}, \citenamefont
  {Calandra}, \citenamefont {Car}, \citenamefont {Cavazzoni}, \citenamefont
  {Ceresoli}, \citenamefont {Chiarotti}, \citenamefont {Cococcioni},
  \citenamefont {Dabo}, \citenamefont {{Dal Corso}}, \citenamefont
  {de~Gironcoli}, \citenamefont {Fabris}, \citenamefont {Fratesi},
  \citenamefont {Gebauer}, \citenamefont {Gerstmann}, \citenamefont
  {Gougoussis}, \citenamefont {Kokalj}, \citenamefont {Lazzeri}, \citenamefont
  {Martin-Samos}, \citenamefont {Marzari}, \citenamefont {Mauri}, \citenamefont
  {Mazzarello}, \citenamefont {Paolini}, \citenamefont {Pasquarello},
  \citenamefont {Paulatto}, \citenamefont {Sbraccia}, \citenamefont {Scandolo},
  \citenamefont {Sclauzero}, \citenamefont {Seitsonen}, \citenamefont
  {Smogunov}, \citenamefont {Umari},\ and\ \citenamefont
  {Wentzcovitch}}]{QE-2009}%
  \BibitemOpen
  \bibfield  {author} {\bibinfo {author} {\bibfnamefont {P.}~\bibnamefont
  {Giannozzi}}, \bibinfo {author} {\bibfnamefont {S.}~\bibnamefont {Baroni}},
  \bibinfo {author} {\bibfnamefont {N.}~\bibnamefont {Bonini}}, \bibinfo
  {author} {\bibfnamefont {M.}~\bibnamefont {Calandra}}, \bibinfo {author}
  {\bibfnamefont {R.}~\bibnamefont {Car}}, \bibinfo {author} {\bibfnamefont
  {C.}~\bibnamefont {Cavazzoni}}, \bibinfo {author} {\bibfnamefont
  {D.}~\bibnamefont {Ceresoli}}, \bibinfo {author} {\bibfnamefont {G.~L.}\
  \bibnamefont {Chiarotti}}, \bibinfo {author} {\bibfnamefont {M.}~\bibnamefont
  {Cococcioni}}, \bibinfo {author} {\bibfnamefont {I.}~\bibnamefont {Dabo}},
  \bibinfo {author} {\bibfnamefont {A.}~\bibnamefont {{Dal Corso}}}, \bibinfo
  {author} {\bibfnamefont {S.}~\bibnamefont {de~Gironcoli}}, \bibinfo {author}
  {\bibfnamefont {S.}~\bibnamefont {Fabris}}, \bibinfo {author} {\bibfnamefont
  {G.}~\bibnamefont {Fratesi}}, \bibinfo {author} {\bibfnamefont
  {R.}~\bibnamefont {Gebauer}}, \bibinfo {author} {\bibfnamefont
  {U.}~\bibnamefont {Gerstmann}}, \bibinfo {author} {\bibfnamefont
  {C.}~\bibnamefont {Gougoussis}}, \bibinfo {author} {\bibfnamefont
  {A.}~\bibnamefont {Kokalj}}, \bibinfo {author} {\bibfnamefont
  {M.}~\bibnamefont {Lazzeri}}, \bibinfo {author} {\bibfnamefont
  {L.}~\bibnamefont {Martin-Samos}}, \bibinfo {author} {\bibfnamefont
  {N.}~\bibnamefont {Marzari}}, \bibinfo {author} {\bibfnamefont
  {F.}~\bibnamefont {Mauri}}, \bibinfo {author} {\bibfnamefont
  {R.}~\bibnamefont {Mazzarello}}, \bibinfo {author} {\bibfnamefont
  {S.}~\bibnamefont {Paolini}}, \bibinfo {author} {\bibfnamefont
  {A.}~\bibnamefont {Pasquarello}}, \bibinfo {author} {\bibfnamefont
  {L.}~\bibnamefont {Paulatto}}, \bibinfo {author} {\bibfnamefont
  {C.}~\bibnamefont {Sbraccia}}, \bibinfo {author} {\bibfnamefont
  {S.}~\bibnamefont {Scandolo}}, \bibinfo {author} {\bibfnamefont
  {G.}~\bibnamefont {Sclauzero}}, \bibinfo {author} {\bibfnamefont {A.~P.}\
  \bibnamefont {Seitsonen}}, \bibinfo {author} {\bibfnamefont {A.}~\bibnamefont
  {Smogunov}}, \bibinfo {author} {\bibfnamefont {P.}~\bibnamefont {Umari}}, \
  and\ \bibinfo {author} {\bibfnamefont {R.~M.}\ \bibnamefont {Wentzcovitch}},\
  }\href@noop {} {\bibfield  {journal} {\bibinfo  {journal} {Journal of
  Physics: Condensed Matter}\ }\textbf {\bibinfo {volume} {21}},\ \bibinfo
  {pages} {395502 (19pp)} (\bibinfo {year} {2009})}\BibitemShut {NoStop}%
\bibitem [{\citenamefont {Ceperley}\ and\ \citenamefont {Alder}(1980)}]{LDA}%
  \BibitemOpen
  \bibfield  {author} {\bibinfo {author} {\bibfnamefont {D.~M.}\ \bibnamefont
  {Ceperley}}\ and\ \bibinfo {author} {\bibfnamefont {B.~J.}\ \bibnamefont
  {Alder}},\ }\href {\doibase 10.1103/PhysRevLett.45.566} {\bibfield  {journal}
  {\bibinfo  {journal} {Phys. Rev. Lett.}\ }\textbf {\bibinfo {volume} {45}},\
  \bibinfo {pages} {566} (\bibinfo {year} {1980})}\BibitemShut {NoStop}%
\bibitem [{\citenamefont {Perdew}\ \emph {et~al.}(1996)\citenamefont {Perdew},
  \citenamefont {Burke},\ and\ \citenamefont
  {Ernzerhof}}]{perdew1996generalized}%
  \BibitemOpen
  \bibfield  {author} {\bibinfo {author} {\bibfnamefont {J.~P.}\ \bibnamefont
  {Perdew}}, \bibinfo {author} {\bibfnamefont {K.}~\bibnamefont {Burke}}, \
  and\ \bibinfo {author} {\bibfnamefont {M.}~\bibnamefont {Ernzerhof}},\
  }\href@noop {} {\bibfield  {journal} {\bibinfo  {journal} {Physical review
  letters}\ }\textbf {\bibinfo {volume} {77}},\ \bibinfo {pages} {3865}
  (\bibinfo {year} {1996})}\BibitemShut {NoStop}%
\bibitem [{\citenamefont {Grimme}(2006)}]{DFT-D2}%
  \BibitemOpen
  \bibfield  {author} {\bibinfo {author} {\bibfnamefont {S.}~\bibnamefont
  {Grimme}},\ }\href@noop {} {\bibfield  {journal} {\bibinfo  {journal}
  {Journal of Computational Chemistry}\ }\textbf {\bibinfo {volume} {27}},\
  \bibinfo {pages} {1787} (\bibinfo {year} {2006})}\BibitemShut {NoStop}%
\bibitem [{\citenamefont {Togo}\ and\ \citenamefont {Tanaka}(2015)}]{phonopy}%
  \BibitemOpen
  \bibfield  {author} {\bibinfo {author} {\bibfnamefont {A.}~\bibnamefont
  {Togo}}\ and\ \bibinfo {author} {\bibfnamefont {I.}~\bibnamefont {Tanaka}},\
  }\href@noop {} {\bibfield  {journal} {\bibinfo  {journal} {Scr. Mater.}\
  }\textbf {\bibinfo {volume} {108}},\ \bibinfo {pages} {1} (\bibinfo {year}
  {2015})}\BibitemShut {NoStop}%
\bibitem [{\citenamefont {Mostofi}\ \emph {et~al.}(2014)\citenamefont
  {Mostofi}, \citenamefont {Yates}, \citenamefont {Pizzi}, \citenamefont {Lee},
  \citenamefont {Souza}, \citenamefont {Vanderbilt},\ and\ \citenamefont
  {Marzari}}]{mostofi2014updated}%
  \BibitemOpen
  \bibfield  {author} {\bibinfo {author} {\bibfnamefont {A.~A.}\ \bibnamefont
  {Mostofi}}, \bibinfo {author} {\bibfnamefont {J.~R.}\ \bibnamefont {Yates}},
  \bibinfo {author} {\bibfnamefont {G.}~\bibnamefont {Pizzi}}, \bibinfo
  {author} {\bibfnamefont {Y.-S.}\ \bibnamefont {Lee}}, \bibinfo {author}
  {\bibfnamefont {I.}~\bibnamefont {Souza}}, \bibinfo {author} {\bibfnamefont
  {D.}~\bibnamefont {Vanderbilt}}, \ and\ \bibinfo {author} {\bibfnamefont
  {N.}~\bibnamefont {Marzari}},\ }\href@noop {} {\bibfield  {journal} {\bibinfo
   {journal} {Computer Physics Communications}\ }\textbf {\bibinfo {volume}
  {185}},\ \bibinfo {pages} {2309} (\bibinfo {year} {2014})}\BibitemShut
  {NoStop}%
\bibitem [{\citenamefont {Guinea}\ \emph {et~al.}(1983)\citenamefont {Guinea},
  \citenamefont {Tejedor}, \citenamefont {Flores},\ and\ \citenamefont
  {Louis}}]{PhysRevB.28.4397}%
  \BibitemOpen
  \bibfield  {author} {\bibinfo {author} {\bibfnamefont {F.}~\bibnamefont
  {Guinea}}, \bibinfo {author} {\bibfnamefont {C.}~\bibnamefont {Tejedor}},
  \bibinfo {author} {\bibfnamefont {F.}~\bibnamefont {Flores}}, \ and\ \bibinfo
  {author} {\bibfnamefont {E.}~\bibnamefont {Louis}},\ }\href {\doibase
  10.1103/PhysRevB.28.4397} {\bibfield  {journal} {\bibinfo  {journal} {Phys.
  Rev. B}\ }\textbf {\bibinfo {volume} {28}},\ \bibinfo {pages} {4397}
  (\bibinfo {year} {1983})}\BibitemShut {NoStop}%
\bibitem [{\citenamefont {Sancho}\ \emph {et~al.}(1984)\citenamefont {Sancho},
  \citenamefont {Sancho},\ and\ \citenamefont {Rubio}}]{Sancho_1984}%
  \BibitemOpen
  \bibfield  {author} {\bibinfo {author} {\bibfnamefont {M.~P.~L.}\
  \bibnamefont {Sancho}}, \bibinfo {author} {\bibfnamefont {J.~M.~L.}\
  \bibnamefont {Sancho}}, \ and\ \bibinfo {author} {\bibfnamefont
  {J.}~\bibnamefont {Rubio}},\ }\href {\doibase 10.1088/0305-4608/14/5/016}
  {\bibfield  {journal} {\bibinfo  {journal} {Journal of Physics F: Metal
  Physics}\ }\textbf {\bibinfo {volume} {14}},\ \bibinfo {pages} {1205}
  (\bibinfo {year} {1984})}\BibitemShut {NoStop}%
\bibitem [{\citenamefont {Sancho}\ \emph {et~al.}(1985)\citenamefont {Sancho},
  \citenamefont {Sancho}, \citenamefont {Sancho},\ and\ \citenamefont
  {Rubio}}]{Sancho_1985}%
  \BibitemOpen
  \bibfield  {author} {\bibinfo {author} {\bibfnamefont {M.~P.~L.}\
  \bibnamefont {Sancho}}, \bibinfo {author} {\bibfnamefont {J.~M.~L.}\
  \bibnamefont {Sancho}}, \bibinfo {author} {\bibfnamefont {J.~M.~L.}\
  \bibnamefont {Sancho}}, \ and\ \bibinfo {author} {\bibfnamefont
  {J.}~\bibnamefont {Rubio}},\ }\href {\doibase 10.1088/0305-4608/15/4/009}
  {\bibfield  {journal} {\bibinfo  {journal} {Journal of Physics F: Metal
  Physics}\ }\textbf {\bibinfo {volume} {15}},\ \bibinfo {pages} {851}
  (\bibinfo {year} {1985})}\BibitemShut {NoStop}%
\bibitem [{\citenamefont {Wu}\ \emph {et~al.}(2018{\natexlab{b}})\citenamefont
  {Wu}, \citenamefont {Zhang}, \citenamefont {Song}, \citenamefont {Troyer},\
  and\ \citenamefont {Soluyanov}}]{WU2017}%
  \BibitemOpen
  \bibfield  {author} {\bibinfo {author} {\bibfnamefont {Q.}~\bibnamefont
  {Wu}}, \bibinfo {author} {\bibfnamefont {S.}~\bibnamefont {Zhang}}, \bibinfo
  {author} {\bibfnamefont {H.-F.}\ \bibnamefont {Song}}, \bibinfo {author}
  {\bibfnamefont {M.}~\bibnamefont {Troyer}}, \ and\ \bibinfo {author}
  {\bibfnamefont {A.~A.}\ \bibnamefont {Soluyanov}},\ }\href {\doibase
  https://doi.org/10.1016/j.cpc.2017.09.033} {\bibfield  {journal} {\bibinfo
  {journal} {Computer Physics Communications}\ }\textbf {\bibinfo {volume}
  {224}},\ \bibinfo {pages} {405 } (\bibinfo {year}
  {2018}{\natexlab{b}})}\BibitemShut {NoStop}%
\bibitem [{\citenamefont {Yu}\ \emph {et~al.}(2011)\citenamefont {Yu},
  \citenamefont {Qi}, \citenamefont {Bernevig}, \citenamefont {Fang},\ and\
  \citenamefont {Dai}}]{Z2_Rui}%
  \BibitemOpen
  \bibfield  {author} {\bibinfo {author} {\bibfnamefont {R.}~\bibnamefont
  {Yu}}, \bibinfo {author} {\bibfnamefont {X.-L.}\ \bibnamefont {Qi}}, \bibinfo
  {author} {\bibfnamefont {A.}~\bibnamefont {Bernevig}}, \bibinfo {author}
  {\bibfnamefont {Z.}~\bibnamefont {Fang}}, \ and\ \bibinfo {author}
  {\bibfnamefont {X.}~\bibnamefont {Dai}},\ }\href {\doibase
  10.1103/PhysRevB.84.075119} {\bibfield  {journal} {\bibinfo  {journal} {Phys.
  Rev. B}\ }\textbf {\bibinfo {volume} {84}} (\bibinfo {year} {2011}),\
  10.1103/PhysRevB.84.075119}\BibitemShut {NoStop}%
\bibitem [{\citenamefont {Soluyanov}\ and\ \citenamefont
  {Vanderbilt}(2011)}]{soluyanov2011computing}%
  \BibitemOpen
  \bibfield  {author} {\bibinfo {author} {\bibfnamefont {A.~A.}\ \bibnamefont
  {Soluyanov}}\ and\ \bibinfo {author} {\bibfnamefont {D.}~\bibnamefont
  {Vanderbilt}},\ }\href@noop {} {\bibfield  {journal} {\bibinfo  {journal}
  {Physical Review B}\ }\textbf {\bibinfo {volume} {83}},\ \bibinfo {pages}
  {235401} (\bibinfo {year} {2011})}\BibitemShut {NoStop}%
\bibitem [{\citenamefont {Sodemann}\ and\ \citenamefont
  {Fu}(2015)}]{PhysRevLett.115.216806}%
  \BibitemOpen
  \bibfield  {author} {\bibinfo {author} {\bibfnamefont {I.}~\bibnamefont
  {Sodemann}}\ and\ \bibinfo {author} {\bibfnamefont {L.}~\bibnamefont {Fu}},\
  }\href {\doibase 10.1103/PhysRevLett.115.216806} {\bibfield  {journal}
  {\bibinfo  {journal} {Phys. Rev. Lett.}\ }\textbf {\bibinfo {volume} {115}},\
  \bibinfo {pages} {216806} (\bibinfo {year} {2015})}\BibitemShut {NoStop}%
\bibitem [{\citenamefont {Tsirkin}(2020)}]{tsirkin2020high}%
  \BibitemOpen
  \bibfield  {author} {\bibinfo {author} {\bibfnamefont {S.~S.}\ \bibnamefont
  {Tsirkin}},\ }\href@noop {} {\enquote {\bibinfo {title} {High performance
  wannier interpolation of berry curvature and related quantities: Wannierberri
  code},}\ } (\bibinfo {year} {2020}),\ \Eprint
  {http://arxiv.org/abs/2008.07992} {arXiv:2008.07992 [cond-mat.mtrl-sci]}
  \BibitemShut {NoStop}%
\bibitem [{\citenamefont {Martin}(2004)}]{martin_2004}%
  \BibitemOpen
  \bibfield  {author} {\bibinfo {author} {\bibfnamefont {R.~M.}\ \bibnamefont
  {Martin}},\ }\enquote {\bibinfo {title} {Functionals for exchange and
  correlation},}\ in\ \href {\doibase 10.1017/CBO9780511805769.010} {\emph
  {\bibinfo {booktitle} {Electronic Structure: Basic Theory and Practical
  Methods}}}\ (\bibinfo  {publisher} {Cambridge University Press},\ \bibinfo
  {year} {2004})\ p.\ \bibinfo {pages} {152–171}\BibitemShut {NoStop}%
\bibitem [{\citenamefont {Xu}\ \emph {et~al.}(2018)\citenamefont {Xu},
  \citenamefont {Ma}, \citenamefont {Shen}, \citenamefont {Fatemi},
  \citenamefont {Wu}, \citenamefont {Chang}, \citenamefont {Chang},
  \citenamefont {Valdivia}, \citenamefont {Chan}, \citenamefont {Gibson} \emph
  {et~al.}}]{xu2018electrically}%
  \BibitemOpen
  \bibfield  {author} {\bibinfo {author} {\bibfnamefont {S.-Y.}\ \bibnamefont
  {Xu}}, \bibinfo {author} {\bibfnamefont {Q.}~\bibnamefont {Ma}}, \bibinfo
  {author} {\bibfnamefont {H.}~\bibnamefont {Shen}}, \bibinfo {author}
  {\bibfnamefont {V.}~\bibnamefont {Fatemi}}, \bibinfo {author} {\bibfnamefont
  {S.}~\bibnamefont {Wu}}, \bibinfo {author} {\bibfnamefont {T.-R.}\
  \bibnamefont {Chang}}, \bibinfo {author} {\bibfnamefont {G.}~\bibnamefont
  {Chang}}, \bibinfo {author} {\bibfnamefont {A.~M.~M.}\ \bibnamefont
  {Valdivia}}, \bibinfo {author} {\bibfnamefont {C.-K.}\ \bibnamefont {Chan}},
  \bibinfo {author} {\bibfnamefont {Q.~D.}\ \bibnamefont {Gibson}},  \emph
  {et~al.},\ }\href@noop {} {\bibfield  {journal} {\bibinfo  {journal} {Nature
  Physics}\ }\textbf {\bibinfo {volume} {14}},\ \bibinfo {pages} {900}
  (\bibinfo {year} {2018})}\BibitemShut {NoStop}%
\bibitem [{\citenamefont {Singh}\ \emph {et~al.}(2020)\citenamefont {Singh},
  \citenamefont {Kim}, \citenamefont {Rabe},\ and\ \citenamefont
  {Vanderbilt}}]{PhysRevLett.125.046402}%
  \BibitemOpen
  \bibfield  {author} {\bibinfo {author} {\bibfnamefont {S.}~\bibnamefont
  {Singh}}, \bibinfo {author} {\bibfnamefont {J.}~\bibnamefont {Kim}}, \bibinfo
  {author} {\bibfnamefont {K.~M.}\ \bibnamefont {Rabe}}, \ and\ \bibinfo
  {author} {\bibfnamefont {D.}~\bibnamefont {Vanderbilt}},\ }\href {\doibase
  10.1103/PhysRevLett.125.046402} {\bibfield  {journal} {\bibinfo  {journal}
  {Phys. Rev. Lett.}\ }\textbf {\bibinfo {volume} {125}},\ \bibinfo {pages}
  {046402} (\bibinfo {year} {2020})}\BibitemShut {NoStop}%
\bibitem [{\citenamefont {Zhang}\ \emph {et~al.}(2018)\citenamefont {Zhang},
  \citenamefont {van~den Brink}, \citenamefont {Felser},\ and\ \citenamefont
  {Yan}}]{zhang2018electrically}%
  \BibitemOpen
  \bibfield  {author} {\bibinfo {author} {\bibfnamefont {Y.}~\bibnamefont
  {Zhang}}, \bibinfo {author} {\bibfnamefont {J.}~\bibnamefont {van~den
  Brink}}, \bibinfo {author} {\bibfnamefont {C.}~\bibnamefont {Felser}}, \ and\
  \bibinfo {author} {\bibfnamefont {B.}~\bibnamefont {Yan}},\ }\href@noop {}
  {\bibfield  {journal} {\bibinfo  {journal} {2D Materials}\ }\textbf {\bibinfo
  {volume} {5}},\ \bibinfo {pages} {044001} (\bibinfo {year}
  {2018})}\BibitemShut {NoStop}%
\bibitem [{\citenamefont {Ma}\ \emph {et~al.}(2019)\citenamefont {Ma},
  \citenamefont {Xu}, \citenamefont {Shen}, \citenamefont {MacNeill},
  \citenamefont {Fatemi}, \citenamefont {Chang}, \citenamefont {Valdivia},
  \citenamefont {Wu}, \citenamefont {Du}, \citenamefont {Hsu} \emph
  {et~al.}}]{ma2019observation}%
  \BibitemOpen
  \bibfield  {author} {\bibinfo {author} {\bibfnamefont {Q.}~\bibnamefont
  {Ma}}, \bibinfo {author} {\bibfnamefont {S.-Y.}\ \bibnamefont {Xu}}, \bibinfo
  {author} {\bibfnamefont {H.}~\bibnamefont {Shen}}, \bibinfo {author}
  {\bibfnamefont {D.}~\bibnamefont {MacNeill}}, \bibinfo {author}
  {\bibfnamefont {V.}~\bibnamefont {Fatemi}}, \bibinfo {author} {\bibfnamefont
  {T.-R.}\ \bibnamefont {Chang}}, \bibinfo {author} {\bibfnamefont {A.~M.~M.}\
  \bibnamefont {Valdivia}}, \bibinfo {author} {\bibfnamefont {S.}~\bibnamefont
  {Wu}}, \bibinfo {author} {\bibfnamefont {Z.}~\bibnamefont {Du}}, \bibinfo
  {author} {\bibfnamefont {C.-H.}\ \bibnamefont {Hsu}},  \emph {et~al.},\
  }\href@noop {} {\bibfield  {journal} {\bibinfo  {journal} {Nature}\ }\textbf
  {\bibinfo {volume} {565}},\ \bibinfo {pages} {337} (\bibinfo {year}
  {2019})}\BibitemShut {NoStop}%
\bibitem [{\citenamefont {Ma}\ \emph {et~al.}(2017)\citenamefont {Ma},
  \citenamefont {Xu}, \citenamefont {Chan}, \citenamefont {Zhang},
  \citenamefont {Chang}, \citenamefont {Lin}, \citenamefont {Xie},
  \citenamefont {Palacios}, \citenamefont {Lin}, \citenamefont {Jia} \emph
  {et~al.}}]{ma2017direct}%
  \BibitemOpen
  \bibfield  {author} {\bibinfo {author} {\bibfnamefont {Q.}~\bibnamefont
  {Ma}}, \bibinfo {author} {\bibfnamefont {S.-Y.}\ \bibnamefont {Xu}}, \bibinfo
  {author} {\bibfnamefont {C.-K.}\ \bibnamefont {Chan}}, \bibinfo {author}
  {\bibfnamefont {C.-L.}\ \bibnamefont {Zhang}}, \bibinfo {author}
  {\bibfnamefont {G.}~\bibnamefont {Chang}}, \bibinfo {author} {\bibfnamefont
  {Y.}~\bibnamefont {Lin}}, \bibinfo {author} {\bibfnamefont {W.}~\bibnamefont
  {Xie}}, \bibinfo {author} {\bibfnamefont {T.}~\bibnamefont {Palacios}},
  \bibinfo {author} {\bibfnamefont {H.}~\bibnamefont {Lin}}, \bibinfo {author}
  {\bibfnamefont {S.}~\bibnamefont {Jia}},  \emph {et~al.},\ }\href@noop {}
  {\bibfield  {journal} {\bibinfo  {journal} {Nature Physics}\ }\textbf
  {\bibinfo {volume} {13}},\ \bibinfo {pages} {842} (\bibinfo {year}
  {2017})}\BibitemShut {NoStop}%
\bibitem [{\citenamefont {Sun}\ \emph {et~al.}(2017)\citenamefont {Sun},
  \citenamefont {Sun}, \citenamefont {Wei}, \citenamefont {Guo}, \citenamefont
  {Tian}, \citenamefont {Chen}, \citenamefont {Yang},\ and\ \citenamefont
  {Li}}]{sun2017circular}%
  \BibitemOpen
  \bibfield  {author} {\bibinfo {author} {\bibfnamefont {K.}~\bibnamefont
  {Sun}}, \bibinfo {author} {\bibfnamefont {S.-S.}\ \bibnamefont {Sun}},
  \bibinfo {author} {\bibfnamefont {L.-L.}\ \bibnamefont {Wei}}, \bibinfo
  {author} {\bibfnamefont {C.}~\bibnamefont {Guo}}, \bibinfo {author}
  {\bibfnamefont {H.-F.}\ \bibnamefont {Tian}}, \bibinfo {author}
  {\bibfnamefont {G.-F.}\ \bibnamefont {Chen}}, \bibinfo {author}
  {\bibfnamefont {H.-X.}\ \bibnamefont {Yang}}, \ and\ \bibinfo {author}
  {\bibfnamefont {J.-Q.}\ \bibnamefont {Li}},\ }\href@noop {} {\bibfield
  {journal} {\bibinfo  {journal} {Chinese Physics Letters}\ }\textbf {\bibinfo
  {volume} {34}},\ \bibinfo {pages} {117203} (\bibinfo {year}
  {2017})}\BibitemShut {NoStop}%
\bibitem [{\citenamefont {Xu}\ \emph {et~al.}(2014)\citenamefont {Xu},
  \citenamefont {Liu}, \citenamefont {Wang}, \citenamefont {Ge}, \citenamefont
  {Liu}, \citenamefont {Yang}, \citenamefont {Chen}, \citenamefont {Liu},
  \citenamefont {Xu}, \citenamefont {Gao} \emph {et~al.}}]{xu2014artificial}%
  \BibitemOpen
  \bibfield  {author} {\bibinfo {author} {\bibfnamefont {J.-P.}\ \bibnamefont
  {Xu}}, \bibinfo {author} {\bibfnamefont {C.}~\bibnamefont {Liu}}, \bibinfo
  {author} {\bibfnamefont {M.-X.}\ \bibnamefont {Wang}}, \bibinfo {author}
  {\bibfnamefont {J.}~\bibnamefont {Ge}}, \bibinfo {author} {\bibfnamefont
  {Z.-L.}\ \bibnamefont {Liu}}, \bibinfo {author} {\bibfnamefont
  {X.}~\bibnamefont {Yang}}, \bibinfo {author} {\bibfnamefont {Y.}~\bibnamefont
  {Chen}}, \bibinfo {author} {\bibfnamefont {Y.}~\bibnamefont {Liu}}, \bibinfo
  {author} {\bibfnamefont {Z.-A.}\ \bibnamefont {Xu}}, \bibinfo {author}
  {\bibfnamefont {C.-L.}\ \bibnamefont {Gao}},  \emph {et~al.},\ }\href@noop {}
  {\bibfield  {journal} {\bibinfo  {journal} {Physical Review Letters}\
  }\textbf {\bibinfo {volume} {112}},\ \bibinfo {pages} {217001} (\bibinfo
  {year} {2014})}\BibitemShut {NoStop}%
\bibitem [{\citenamefont {Kim}\ \emph {et~al.}(2017)\citenamefont {Kim},
  \citenamefont {Park}, \citenamefont {Lee}, \citenamefont {Lee}, \citenamefont
  {Park}, \citenamefont {Lee}, \citenamefont {Lee},\ and\ \citenamefont
  {Lee}}]{kim2017strong}%
  \BibitemOpen
  \bibfield  {author} {\bibinfo {author} {\bibfnamefont {M.}~\bibnamefont
  {Kim}}, \bibinfo {author} {\bibfnamefont {G.-H.}\ \bibnamefont {Park}},
  \bibinfo {author} {\bibfnamefont {J.}~\bibnamefont {Lee}}, \bibinfo {author}
  {\bibfnamefont {J.~H.}\ \bibnamefont {Lee}}, \bibinfo {author} {\bibfnamefont
  {J.}~\bibnamefont {Park}}, \bibinfo {author} {\bibfnamefont {H.}~\bibnamefont
  {Lee}}, \bibinfo {author} {\bibfnamefont {G.-H.}\ \bibnamefont {Lee}}, \ and\
  \bibinfo {author} {\bibfnamefont {H.-J.}\ \bibnamefont {Lee}},\ }\href@noop
  {} {\bibfield  {journal} {\bibinfo  {journal} {Nano Letters}\ }\textbf
  {\bibinfo {volume} {17}},\ \bibinfo {pages} {6125} (\bibinfo {year}
  {2017})}\BibitemShut {NoStop}%
\bibitem [{\citenamefont {Huang}\ \emph {et~al.}(2018)\citenamefont {Huang},
  \citenamefont {Narayan}, \citenamefont {Zhang}, \citenamefont {Liu},
  \citenamefont {Yan}, \citenamefont {Wang}, \citenamefont {Zhang},
  \citenamefont {Wang}, \citenamefont {Zhou}, \citenamefont {Yi} \emph
  {et~al.}}]{huang2018inducing}%
  \BibitemOpen
  \bibfield  {author} {\bibinfo {author} {\bibfnamefont {C.}~\bibnamefont
  {Huang}}, \bibinfo {author} {\bibfnamefont {A.}~\bibnamefont {Narayan}},
  \bibinfo {author} {\bibfnamefont {E.}~\bibnamefont {Zhang}}, \bibinfo
  {author} {\bibfnamefont {Y.}~\bibnamefont {Liu}}, \bibinfo {author}
  {\bibfnamefont {X.}~\bibnamefont {Yan}}, \bibinfo {author} {\bibfnamefont
  {J.}~\bibnamefont {Wang}}, \bibinfo {author} {\bibfnamefont {C.}~\bibnamefont
  {Zhang}}, \bibinfo {author} {\bibfnamefont {W.}~\bibnamefont {Wang}},
  \bibinfo {author} {\bibfnamefont {T.}~\bibnamefont {Zhou}}, \bibinfo {author}
  {\bibfnamefont {C.}~\bibnamefont {Yi}},  \emph {et~al.},\ }\href@noop {}
  {\bibfield  {journal} {\bibinfo  {journal} {ACS nano}\ }\textbf {\bibinfo
  {volume} {12}},\ \bibinfo {pages} {7185} (\bibinfo {year}
  {2018})}\BibitemShut {NoStop}%
\bibitem [{\citenamefont {Huang}\ \emph {et~al.}(2020)\citenamefont {Huang},
  \citenamefont {Narayan}, \citenamefont {Zhang}, \citenamefont {Xie},
  \citenamefont {Ai}, \citenamefont {Liu}, \citenamefont {Yi}, \citenamefont
  {Shi}, \citenamefont {Sanvito},\ and\ \citenamefont {Xiu}}]{huang2020edge}%
  \BibitemOpen
  \bibfield  {author} {\bibinfo {author} {\bibfnamefont {C.}~\bibnamefont
  {Huang}}, \bibinfo {author} {\bibfnamefont {A.}~\bibnamefont {Narayan}},
  \bibinfo {author} {\bibfnamefont {E.}~\bibnamefont {Zhang}}, \bibinfo
  {author} {\bibfnamefont {X.}~\bibnamefont {Xie}}, \bibinfo {author}
  {\bibfnamefont {L.}~\bibnamefont {Ai}}, \bibinfo {author} {\bibfnamefont
  {S.}~\bibnamefont {Liu}}, \bibinfo {author} {\bibfnamefont {C.}~\bibnamefont
  {Yi}}, \bibinfo {author} {\bibfnamefont {Y.}~\bibnamefont {Shi}}, \bibinfo
  {author} {\bibfnamefont {S.}~\bibnamefont {Sanvito}}, \ and\ \bibinfo
  {author} {\bibfnamefont {F.}~\bibnamefont {Xiu}},\ }\href@noop {} {\bibfield
  {journal} {\bibinfo  {journal} {National Science Review}\ } (\bibinfo {year}
  {2020})}\BibitemShut {NoStop}%
\bibitem [{\citenamefont {Trainer}\ \emph {et~al.}(2020)\citenamefont
  {Trainer}, \citenamefont {Wang}, \citenamefont {Bobba}, \citenamefont
  {Samuelson}, \citenamefont {Xi}, \citenamefont {Zasadzinski}, \citenamefont
  {Nieminen}, \citenamefont {Bansil},\ and\ \citenamefont
  {Iavarone}}]{trainer2020proximity}%
  \BibitemOpen
  \bibfield  {author} {\bibinfo {author} {\bibfnamefont {D.~J.}\ \bibnamefont
  {Trainer}}, \bibinfo {author} {\bibfnamefont {B.}~\bibnamefont {Wang}},
  \bibinfo {author} {\bibfnamefont {F.}~\bibnamefont {Bobba}}, \bibinfo
  {author} {\bibfnamefont {N.}~\bibnamefont {Samuelson}}, \bibinfo {author}
  {\bibfnamefont {X.}~\bibnamefont {Xi}}, \bibinfo {author} {\bibfnamefont
  {J.}~\bibnamefont {Zasadzinski}}, \bibinfo {author} {\bibfnamefont
  {J.}~\bibnamefont {Nieminen}}, \bibinfo {author} {\bibfnamefont
  {A.}~\bibnamefont {Bansil}}, \ and\ \bibinfo {author} {\bibfnamefont
  {M.}~\bibnamefont {Iavarone}},\ }\href@noop {} {\bibfield  {journal}
  {\bibinfo  {journal} {ACS nano}\ }\textbf {\bibinfo {volume} {14}},\ \bibinfo
  {pages} {2718} (\bibinfo {year} {2020})}\BibitemShut {NoStop}%
\bibitem [{\citenamefont {L{\"u}pke}\ \emph {et~al.}(2020)\citenamefont
  {L{\"u}pke}, \citenamefont {Waters}, \citenamefont {Sergio}, \citenamefont
  {Widom}, \citenamefont {Mandrus}, \citenamefont {Yan}, \citenamefont
  {Feenstra},\ and\ \citenamefont {Hunt}}]{lupke2020proximity}%
  \BibitemOpen
  \bibfield  {author} {\bibinfo {author} {\bibfnamefont {F.}~\bibnamefont
  {L{\"u}pke}}, \bibinfo {author} {\bibfnamefont {D.}~\bibnamefont {Waters}},
  \bibinfo {author} {\bibfnamefont {C.}~\bibnamefont {Sergio}}, \bibinfo
  {author} {\bibfnamefont {M.}~\bibnamefont {Widom}}, \bibinfo {author}
  {\bibfnamefont {D.~G.}\ \bibnamefont {Mandrus}}, \bibinfo {author}
  {\bibfnamefont {J.}~\bibnamefont {Yan}}, \bibinfo {author} {\bibfnamefont
  {R.~M.}\ \bibnamefont {Feenstra}}, \ and\ \bibinfo {author} {\bibfnamefont
  {B.~M.}\ \bibnamefont {Hunt}},\ }\href@noop {} {\bibfield  {journal}
  {\bibinfo  {journal} {Nature Physics}\ }\textbf {\bibinfo {volume} {16}},\
  \bibinfo {pages} {526} (\bibinfo {year} {2020})}\BibitemShut {NoStop}%
\end{thebibliography}

%

\end{document}